\documentclass[10pt,aps,twocolumn,prd,superscriptaddress,showpacs,nofootinbib,noshowkeys,floatfix,preprintnumbers]{revtex4-1}
\usepackage{graphics,graphicx}
\usepackage{amsmath, amssymb}
\usepackage{multirow}
\usepackage{color}
\usepackage{verbatim}
\usepackage{hyperref}
\usepackage[normalem]{ulem}  
\usepackage{ulem}
\usepackage{color}
\usepackage{cancel}
\usepackage{mathtools}
\usepackage{epstopdf}

\renewcommand\sout{\bgroup \color{blue} \ULdepth=-.5ex \ULset}
\def\slashchar#1{\setbox0=\hbox{$#1$}  
\dimen0=\wd0     
\setbox1=\hbox{/} \dimen1=\wd1  
\ifdim\dimen0>\dimen1   
\rlap{\hbox to \dimen0{\hfil/\hfil}} 
#1     
\else     
\rlap{\hbox to \dimen1{\hfil$#1$\hfil}} 
/      
\fi}

\newcommand{\dd}{\mathrm{d}}

\begin{document}

   \title{Chiral symmetry restoration by parity doubling and the structure of neutron stars}
   \date{\today}
   \author{Micha\l{} Marczenko}
   \affiliation{Institute of Theoretical Physics, University of Wroc\l{}aw, PL-50204 Wroc\l{}aw, Poland}
   \author{David Blaschke}
   \affiliation{Institute of Theoretical Physics, University of Wroc\l{}aw, PL-50204 Wroc\l{}aw, Poland}
   \affiliation{Bogoliubov Laboratory of Theoretical Physics, Joint Institute for Nuclear Research, 141980 Dubna, Russia}
   \affiliation{National Research Nuclear University, 115409 Moscow, Russia}
   \author{Krzysztof Redlich}
   \affiliation{Institute of Theoretical Physics, University of Wroc\l{}aw, PL-50204 Wroc\l{}aw, Poland}
   \affiliation{Extreme Matter Institute EMMI, GSI, D-64291 Darmstadt, Germany}
   \author{Chihiro Sasaki}
   \affiliation{Institute of Theoretical Physics, University of Wroc\l{}aw, PL-50204 Wroc\l{}aw, Poland}

\begin{abstract}
   We investigate the equation of state for the recently developed hybrid quark-meson-nucleon model under neutron star conditions of $\beta-$equilibrium and charge neutrality. The model has the characteristic feature that, at increasing baryon density, the chiral symmetry is restored within the hadronic phase by lifting the mass splitting between chiral partner states, before quark deconfinement takes place. Most important for this study are the nucleon (neutron, proton) and $N(1535)$ states. We present different sets for two free parameters, which result in compact star \mbox{mass-radius} relations in accordance with modern constraints on the mass from PSR~J0348+0432 and on the compactness from GW170817. We also consider the threshold for the direct URCA process for which a new relationship is given, and suggest as an additional constraint on the parameter choice of the model that this process shall become operative at best for stars with masses above the range for binary radio pulsars, $M>1.4~M_\odot$.
\end{abstract}
\maketitle 

\section{Introduction}
\label{sec:introduction}
   
   The investigation of the equation of state~(EoS) of compact star matter became rather topical within the past few years, mainly due to the one-to-one correspondence between EoS and \mbox{mass-radius}~(M-R) relationship~\cite{Lindblom:1998} for the corresponding sequence of compact stars via the solution of the Tolman-Oppenheimer-Volkoff (TOV) equations~\cite{Tolman:1939jz,Oppenheimer:1939ne}. Masses and radii of pulsars are target of observational programs, which can therefore provide stringent constraints to the EoS and phase structure of quantum chromodynamics (QCD) in a region of the QCD phase diagram that is inaccessible to terrestrial experiments and present techniques of lattice QCD simulations. For the extraction of the compact star EoS via Bayesian analysis techniques using mass and radius measurements as priors see Refs.~\cite{Steiner:2010fz,Steiner:2012xt,Alvarez-Castillo:2016oln}. In particular, in the era of multi-messenger astronomy, it shall soon become possible to constrain the sequence of stable compact star configurations in the \mbox{mass-radius} plane inasmuch that a benchmark for the EoS of cold and dense matter can be deduced from it. This would play a similar role for the development of effective models of the QCD phase structure at low temperature and high density, as the ab-initio calculations of lattice QCD at finite temperature and vanishing baryon density~\cite{Bazavov:2014pvz}.
   
   In the study of cold and dense QCD and its applications, commonly used are separate effective models for the nuclear and quark matter phases (two-phase approaches) with a priori assumed first-order phase transition, typically associated with simultaneous chiral and deconfinement transitions. Within this setting, for a constant-speed-of-sound (CSS) model of high-density (quark) matter, a systematic classification of hybrid compact star solutions has been given in~\cite{Alford:2013aca}, which gives a possibility to identify a strong first-order transition in the EoS by the fact that the hybrid star branch in the M-R diagram becomes disconnected from the branch of pure neutron stars. However, already before this occurs, a strong phase transition manifests itself by the appearance of an almost horizontal branch on which the hybrid star solutions lie, as opposed to the merely vertical branch of pure neutron stars. In the literature, this strong phase transition has been discussed as due to quark deconfinement~\cite{Alvarez-Castillo:2016wqj}. This conclusion, as we shall demonstrate in this work, may however be premature since strong phase transitions with a large latent heat occur also within hadronic matter, for instance due to the onset of $\Delta-$matter~\cite{Kolomeitsev:2016ptu} or just the chiral restoration transition inside the nuclear matter phase, which we discuss here.
   
   In this work, we explore the implications of dynamical sequential phase transitions at high baryon density on the structure of neutron stars. To this end, we employ the hybrid \mbox{quark-meson-nucleon} (QMN) model~\cite{Benic:2015pia, Marczenko:2017huu} and extend it by including the isovector $\rho$ meson. We demonstrate how \mbox{high-mass} star configurations can be achieved with different neutron-star interior. We find that, depending on the parametrization, the model predicts different stable configurations with similar mass $M\simeq2~M_\odot$. Our main focus is on the role of the chiral symmetry restoration in the \mbox{high-mass} part of the \mbox{mass-radius} sequence.
   
   In the model parametrization, we pay special attention to recent observational constraints from neutron star physics, in particular from the precise measurement of the high mass $2.01(4)~M_\odot$ for PSR~J0348+432~\cite{Antoniadis:2013pzd}, and for the compactness constraint at $M=1.4~M_\odot$ from the gravitational wave detection of the inspiral phase of the binary neutron star merger GW170817~\cite{TheLIGOScientific:2017qsa}, which constrains the radius at this mass to $13.6~$km~\cite{Annala:2017llu}. Furthermore, we consider the so-called "direct URCA constraint" for the cooling of compact stars which gives a constraint on the admissible proton fraction in typical-mass compact stars (see, e.g., Ref.~\cite{Popov:2005xa}).
   
   This paper is organized as follows. In Sec.~\ref{sec:hqmn_model}, we introduce the hybrid quark-meson-nucleon model, as well as its extension to arbitrary isospin. In Sec.~\ref{sec:eos}, we discuss the obtained numerical results on the equation of state under the neutron-star conditions. In Sec.~\ref{sec:results}, we discuss the obtained neutron-star relations, the direct URCA process, the tidal deformability, as well as possible realizations of the low-temperature phase diagram. Finally, Sec.~\ref{sec:conclusion} is devoted to summary and conclusions.
   
\section{Hybrid Quark-Meson-Nucleon model}
\label{sec:hqmn_model}
   
   \begin{table*}[t!]
   \begin{center}
   \begin{tabular}{|c||c|c|c|c|c|c|c|c|c|}
      \hline
      $m_0$~[MeV] & $m_\sigma$~[MeV] & $g_\omega$ &  $g_\rho$ & $g_1$ & $g_2$ & $g_q$ & $\kappa_b~$[MeV] & $\lambda_b$ \\ \hline\hline
      790         & 370.58           & 6.80       & 7.98      & \multirow{4}{*}{13.00} & \multirow{4}{*}{6.96} & \multirow{4}{*}{3.36} & \multirow{4}{*}{155} & \multirow{4}{*}{0.074} \\ \cline{1-4}
      800         & 363.68           & 6.54       & 8.00      & \multirow{4}{*}{} & \multirow{4}{*}{} & \multirow{4}{*}{} & \multirow{4}{*}{} & \multirow{4}{*}{} \\ \cline{1-4}
      820         & 348.68           & 6.00       & 8.06      & \multirow{4}{*}{} & \multirow{4}{*}{} & \multirow{4}{*}{} & \multirow{4}{*}{} & \multirow{4}{*}{} \\ \cline{1-4}
      840         & 313.83           & 5.44       & 8.11      & \multirow{4}{*}{} & \multirow{4}{*}{} & \multirow{4}{*}{} & \multirow{4}{*}{} & \multirow{4}{*}{} \\ \hline
   \end{tabular}
   \end{center}
   \caption{Sets of the model parameters used in this work. The values of $m_\sigma$ and $g_\omega$ are fixed by the nuclear groundstate properties (see Ref.~\cite{Marczenko:2017huu}), while $g_\rho$ by the symmetry energy (see text). The remaining parameters do not depend on the choice of $m_0$, and their values are taken from Ref.~\cite{Marczenko:2017huu}}
   \label{tab:model_params}
   \end{table*}
   
   In this section, we briefly introduce the hybrid quark-meson-nucleon (QMN) model for the QCD transitions at finite temperature and density~\cite{Benic:2015pia, Marczenko:2017huu} for the \mbox{isospin-symmetric} matter, and extend it to an arbitrary isospin for the application to the physics of neutron stars.
   
   The hybrid QMN model is composed of the baryonic parity-doublers~\cite{Detar:1988kn,Jido:1999hd,Jido:2001nt} and mesons as in the Walecka model, as well as quark degrees of freedom as in the standard quark-meson model. The spontaneous chiral symmetry breaking yields the mass splitting between the two baryonic parity partners, while it generates entire mass of a quark. In this work, we consider a system with $N_f=2$, hence, relevant for this study are the lowest nucleons and their chiral partners, as well as the up and down quarks. The hadronic degrees of freedom are coupled to the chiral fields $\left(\sigma, \boldsymbol\pi\right)$, and the iso-singlet vector field $\omega_\mu$, while the quarks are coupled solely to the former. Another important feature of the hybrid QMN model is that it realizes the concept of statistical confinement by taking into account a medium-dependent modification of the distribution functions, where an additional real scalar field $b$ is introduced.
   
   Asymptotic freedom in QCD suggests that active degrees of freedom are different depending on their momenta. This naturally supports to introduce a UV cutoff for the hadrons and IR cutoff for the quarks. This notion has been widely used in effective theories and Dyson-Schwinger approaches~\cite{Roberts:2010rn, Roberts:2011wy}. As demonstrated in~\cite{Benic:2015pia}, the expectation value of the $b$ field plays a role similar to that of the Polyakov loop at finite temperature and vanishing $\mu_q$, and it is responsible for the onset of quark degrees of freedom around the crossover. Therefore, the $b$ field is likely related with gluons, especially the chromo-electric component. So far it is not known how one is transmuted to another since there is no rigorous order parameter of deconfinement in QCD with light fermions, but the introduction of the $b$ field is a practical and minimal way to realize the physical situation where quarks are not activated deep in the hadronic phase.
   
   In the mean-field approximation, the thermodynamic potential of the hybrid QMN model reads\footnote{We note that the original hybrid QMN model deals with conformal anomaly as well~\cite{Benic:2015pia}. However, the presence of the dilaton (glueball) does not affect the nuclear groundstate properties in the mean-field approximation due to its heavy mass. Moreover, its expectation value is almost constant in the region of our interest. Hence, following our previous work~\cite{Marczenko:2017huu}, we do not take the dilaton dynamics into account.}
   \begin{equation}\label{eq:thermo_pot}
      \Omega = \sum_{x}\Omega_x + V_\sigma + V_\omega + V_b \textrm,
   \end{equation}
   where the summation in the first term goes over the up~($u$) and down~($d$) quarks, as well as the nucleonic states with positive and negative parity. The positive-parity nucleons correspond to the positively charged and neutral $N(938)$ states, i.e., proton ($p_+$) and neutron ($n_+$), respectively. The \mbox{negative-parity} nucleons are identified as their counterparts, $N(1535)$~\cite{Patrignani:2016xqp}, and are denoted as $p_-$ and $n_-$. The kinetic term, $\Omega_x$, reads
   \begin{equation}
      \Omega_x = \gamma_x \int\frac{\dd^3p}{\left(2\pi\right)^3} T \left[\ln\left(1-n_x\right) + \ln\left(1-\bar n_x\right)\right]\textrm.
   \end{equation}
   The factor \mbox{$\gamma_\pm=2$} denotes the spin degeneracy for the nucleons with positive/negative parity, and \mbox{$\gamma_q=2\times 3 = 6$} is the spin-color degeneracy factor for the quarks. The functions $n_x$ are the modified \mbox{Fermi-Dirac} distribution functions for the nucleons
   \begin{equation}\label{eq:cutoff_nuc}
   \begin{split}
      n_\pm &= \theta \left(\alpha^2 b^2 - \boldsymbol p^2\right) f_\pm \textrm,\\
      \bar n_\pm &= \theta \left(\alpha^2 b^2 - \boldsymbol p^2\right) \bar f_\pm \textrm,
   \end{split}
   \end{equation}
   and for the quarks, accordingly
   \begin{equation}\label{eq:cutoff_quark}
   \begin{split}
      n_q &= \theta \left(\boldsymbol p^2-b^2\right) f_q \textrm,\\
      \bar n_q &= \theta \left(\boldsymbol p^2-b^2\right) \bar f_q \textrm,
   \end{split}
   \end{equation}
   where $b$ is the expectation value of the $b$-field, and $\alpha$ is a dimensionless model parameter~\cite{Benic:2015pia, Marczenko:2017huu}. From the definition of $n_\pm$ and $n_q$, it is evident that, to mimic the statistical confinement, the expected behavior of the $b$ field is to have a non-trivial vacuum expectation value, in order to favor the hadronic degrees of freedom over the quark ones at low densities. On the other hand, it is expected that it vanishes at higher densities in order to suppress the hadronic degrees of freedom and to allow for the population of quarks. This is achieved by allowing $b$ to be generated from a potential $V_b$ (to be introduced later in this section).
   
   The functions $f_x$ and $\bar f_x$ are the standard Fermi-Dirac distributions,
   \begin{equation}
   \begin{split}
      f_x &= \frac{1}{1+e^{\beta \left(E_x - \mu_x\right)}} \textrm,\\
      \bar f_x &= \frac{1}{1+e^{\beta \left(E_x + \mu_x\right)}}\textrm,
   \end{split}
   \end{equation}
   with $\beta$ being the inverse temperature, the dispersion relation $E_x = \sqrt{\boldsymbol p^2 + m_x^2}$. The effective chemical potentials for $p_\pm$ and $n_\pm$ are defined as\footnote{In the mean-field approximation, the non-vanishing expectation value of the $\omega$ field is the time-like component, hence we simply denote it by $\omega_0 \equiv \omega$.}
   \begin{equation}\label{eq:u_eff_had}
   \begin{split}
      \mu_{p_\pm} &= \mu_B - g_\omega\omega + \mu_Q\textrm,\\
      \mu_{n_\pm} &= \mu_B - g_\omega\omega \textrm.
   \end{split}
   \end{equation}
   The constant $g_\omega$ couples the nucleons to the $\omega$ field. Its strength is fixed by the nuclear saturation properties~\cite{Benic:2015pia,Marczenko:2017huu}. The effective chemical potentials for up and down quarks are given by
   \begin{equation}\label{eq:u_effq}
   \begin{split}
      \mu_u &= \frac{1}{3}\mu_B + \frac{2}{3}\mu_Q\textrm,\\
      \mu_d &= \frac{1}{3}\mu_B - \frac{1}{3}\mu_Q\textrm.
   \end{split}
   \end{equation}
   In Eqs.~(\ref{eq:u_eff_had})~and~(\ref{eq:u_effq}), $\mu_B$, $\mu_Q$ are the baryon and charge chemical potentials, respectively. The effective masses of the parity doublers \mbox{$m_{p_\pm} = m_{n_\pm} \equiv m_\pm$}, are given by
   \begin{equation}\label{eq:mass_had}
      m_\pm = \frac{1}{2}\left[\sqrt{\left(g_1+g_2\right)^2\sigma^2 + 4m_0^2} \mp \left( g_1 - g_2 \right) \sigma \right] \textrm,
   \end{equation}
   and for quarks, $m_u = m_d \equiv m_q$,
   \begin{equation}\label{eq:mass_quark}
      m_q = g_\sigma \sigma \textrm.
   \end{equation}
   The parameters $g_1$, $g_2$, $g_\sigma$ are Yukawa-coupling constants, $m_0$ is the chirally invariant mass of the baryons and is treated as an external parameter (for more details, see Ref.~\cite{Marczenko:2017huu}). The values of those couplings can be determined by fixing the fermion masses. We take the vacuum masses of the positive- and negative-parity hadronic states to be \mbox{$m_+=939~$MeV}, \mbox{$m_-=1500~$MeV}, respectively. The quark mass is assumed to be $m_+ = 3 m_q$ in the vacuum. When the chiral symmetry is restored, the masses of the baryonic parity partners become degenerate with a common finite mass $m_\pm\left(\sigma=0\right) = m_0$, which reflects the parity doubling structure of the \mbox{low-lying} baryons. This is in contrast to the quarks, which become massless as the chiral symmetry gets restored.
   
   The potentials in Eq.~(\ref{eq:thermo_pot}) are as in the ordinary SU(2) linear sigma model,
   \begin{subequations}\label{eq:potentials}
   \begin{align}
      V_\sigma &= -\frac{\lambda_2}{2}\left(\sigma^2 + \boldsymbol\pi^2\right) + \frac{\lambda_4}{4}\left(\sigma^2 + \boldsymbol\pi^2\right)^2 - \epsilon\sigma \textrm,\\
      V_\omega &= -\frac{1}{2}m_\omega^2 \omega_\mu\omega^\mu\textrm,\\
      V_b &= -\frac{1}{2} \kappa_b^2 b^2 + \frac{1}{4}\lambda_b b^4 \textrm,
   \end{align}
   \end{subequations}
   where the parameters $\lambda_2$, $\lambda_4$, and $\epsilon$ are
   \begin{equation}\label{eq:parity_params}
      \lambda_2 = \frac{m_\sigma^2 - 3 m_\pi^2}{2} \textrm{, }\;\;\;\; \lambda_4 = \frac{ m_\sigma^2 - m_\pi^2 }{2f_\pi^2}\textrm{, }\;\;\;\; \epsilon = m_\pi^2 f_\pi \textrm,
   \end{equation}
   where the pion mass $m_\pi=138~$MeV, pion decay constant $f_\pi=93~$MeV. The sigma mass, $m_\sigma$, is treated as an external parameter, and is fixed by the properties of the nuclear groundstate. The constants $\kappa_b$ and $\lambda_b$ are fixed following Ref.~\cite{Benic:2015pia}. 
   
   In this work, in order to study matter under neutron star conditions, we extend the hybrid QMN model to the case of arbitrary isospin asymmetry. To this end, we introduce an additional term to the Lagrangian density for symmetric matter together with a Yukawa coupling to the isovector $\rho$ meson field~\cite{glendenning00:book}, namely
   \begin{equation}\label{eq:rho_lagrangian}
      \mathcal{L}_\rho = -\frac{1}{2}g_\rho\sum_{k=1,2}\bar\psi_k \boldsymbol\tau \cdot \slashchar{\boldsymbol\rho} \psi_k - \frac{1}{4} {\boldsymbol\rho}_{\mu\nu} \cdot {\boldsymbol \rho}^{\mu\nu} - V_\rho \textrm,
   \end{equation}
   where $\psi_k$ is a set of the the baryonic chiral fields. The term ${\boldsymbol \rho}^{\mu\nu} = \partial^\mu {\boldsymbol \rho}^\nu - \partial^\nu{\boldsymbol \rho}^\mu - g_\rho{\boldsymbol\rho}^\mu \times {\boldsymbol \rho}^\nu$ is the \mbox{field-strength} tensor of the isovector field, and the mass term
   \begin{equation}
      V_\rho = - \frac{1}{2}m_\rho^2{\boldsymbol \rho}_\mu{\boldsymbol \rho}^\mu \textrm,
   \end{equation}
   with $m_\rho$ being the mass of the $\rho$ meson.
   
   Taking the extension from Eq.~(\ref{eq:rho_lagrangian}) into account, the mean-field thermodynamic potential is obtained as
   \begin{equation}\label{eq:thermo_pot_iso}
      \Omega = \sum_{x}\Omega_x + V_\sigma + V_\omega + V_b + V_\rho \textrm.
   \end{equation}
   Due to the inclusion of the $\rho$ meson, the effective chemical potentials~(\ref{eq:u_eff_had}) now become\footnote{
   We denote the non-vanishing component, time-like and neutral, by $\rho_{03} \equiv \rho $, just like the $\omega$.},
   \begin{equation}\label{eq:u_eff_had_iso}
   \begin{split}
      \mu_{p_\pm} &= \mu_B - g_\omega\omega - \frac{1}{2}g_\rho \rho + \mu_Q\textrm,\\
      \mu_{n_\pm} &= \mu_B - g_\omega\omega + \frac{1}{2}g_\rho \rho\textrm.
   \end{split}
   \end{equation}
   
   In the quark sector, a coupling of quarks to the $\rho$ mean field in principle can also be introduced. In order to avoid unnecessary complexity and to reduce the number of free parameters, we do not take it into account in the current work. As a result of this assumption, the effective chemical potentials for quarks given in Eq.~(\ref{eq:u_effq}) remain unaffected.
   
   In the grand canonical ensemble, the net-baryon number density for a species $x$ is defined as
   \begin{equation}
      \rho^x_B = - \frac{\partial \Omega_x}{\partial \mu_B}
   \end{equation}
   The total net-baryon number density is then the sum over all species, namely
   \begin{equation}
      \rho_B = \rho_B^{p_+} + \rho_B^{p_-} + \rho_B^{n_+} + \rho_B^{n_-} + \rho_B^{u} + \rho_B^{d} \textrm.
   \end{equation}
   The particle-density fractions are defined as 
   \begin{equation}
      Y_x = \frac{\rho_B^x}{\rho_B} \textrm.
   \end{equation}
   
   \begin{figure}[t!]
   \begin{center}
      \includegraphics[width=\linewidth]{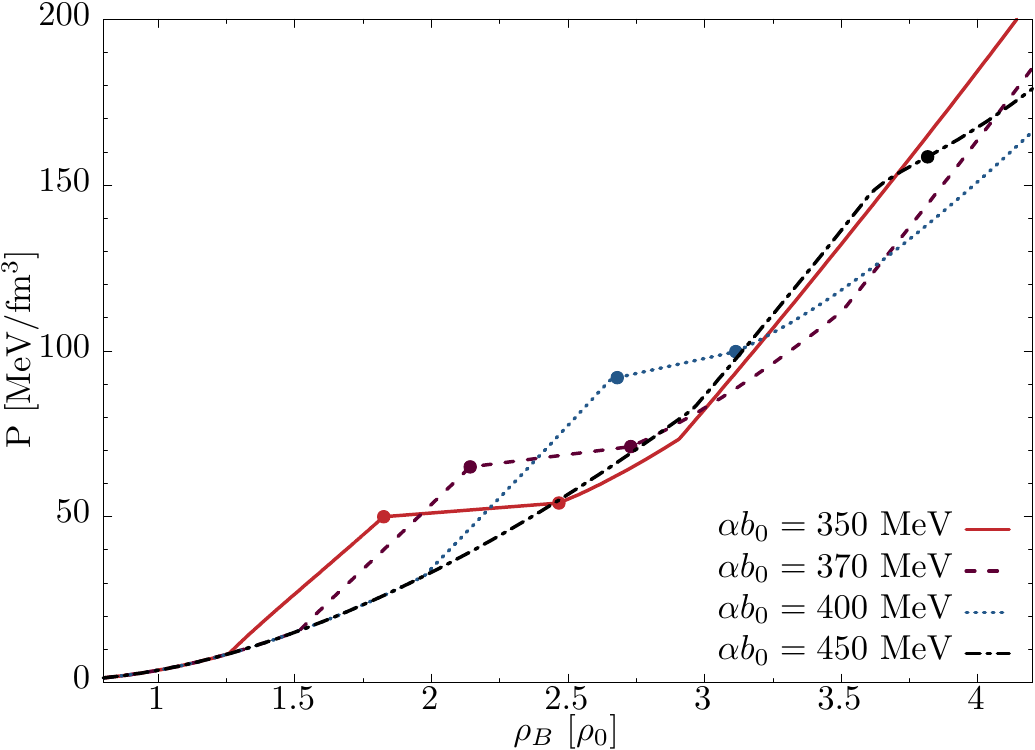}
      \caption{Thermodynamic pressure $P$ as a function of the net-baryon number density $\rho_B$, in units of the saturation density, $\rho_0=0.16$~fm$^{-3}$ for $m_0=790~$MeV. The regions between circles correspond to the coexistence of chirally broken and restored phases in the first-order phase transition. For \mbox{$\alpha b_0=450~$MeV} the transition is a crossover. The deconfinement transitions are triggered at higher densities and are not shown here.}
      \label{fig:eos}
   \end{center}
   \end{figure}
   
   \begin{figure}[t!]
   \begin{center}
      \includegraphics[width=\linewidth]{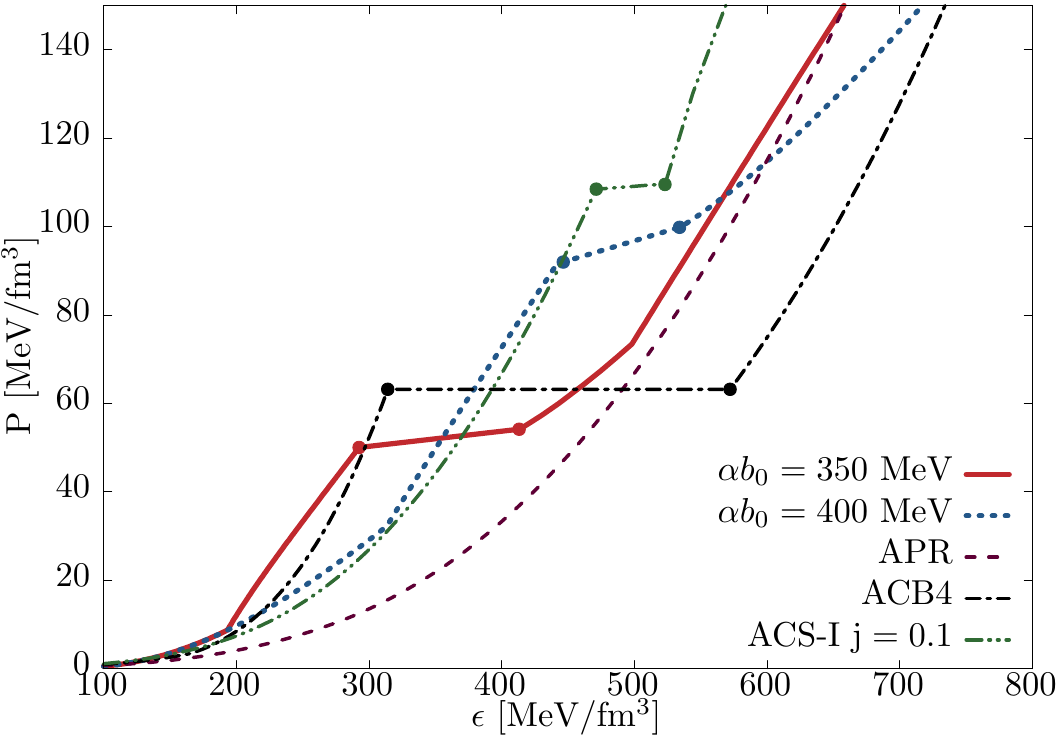}
      \caption{Thermodynamic pressure $P$ as a function of the energy density $\epsilon$ for two parametrizations of the hybrid QMN model for $m_0=790~$MeV. For comparison, shown are also three EoS from Ref.~\cite{Paschalidis:2017qmb}: (i) the APR EoS (purple dashed line) (ii) the \mbox{multi-polytrope} EoS labeled ACB4 (black \mbox line), as well as (iii) the hybrid EoS ACS-I (green \mbox{dash-doubly-dotted} line).}
      \label{fig:eos-comp}
   \end{center}
   \end{figure}
   
   The inclusion of the isovector field as in Eq.~(\ref{eq:rho_lagrangian}) introduces two additional parameters, i.e., the vacuum mass of the $\rho$ meson, $m_\rho$ and the coupling, $g_\rho$. The former is set to be $m_\rho = 775$~MeV~\cite{Patrignani:2016xqp}, while the value of the latter can be fixed by fitting the value of symmetry energy, \mbox{$E_{\textrm{sym}}=31$~MeV}~\cite{Lattimer:2012xj}. The precise strength of the coupling depends on the value of the chirally invariant mass $m_0$. Following the previous studies of the \mbox{parity-doublet-based} models~\cite{Zschiesche:2006zj, Benic:2015pia, Marczenko:2017huu}, as well as recent lattice QCD results~\cite{Aarts:2015mma,Aarts:2017rrl}, we choose rather large values, $m_0=790,~800,~820,~840$~MeV. 
   
   Also, by imposing the empirical values of the binding energy, and the saturation density at zero temperature, 
   
   \begin{equation}
   \begin{split}
      E/A(\mu_B=923~{\rm MeV})- m_+ &= -16~\rm MeV \textrm,\\
      \rho_B(\mu_B=923~{\rm MeV}) &= 0.16~\rm fm^{-3} \textrm,
   \end{split}
   \end{equation}
   and the Stefan-Boltzmann limit at asymptotically high density, the allowed range for the parameter $\alpha$ is set to be $\alpha b_0 = 300 -450~$MeV~\cite{Marczenko:2017huu}. In the present work, we take four representative values within that interval, namely $\alpha b_0 = 350,~370,~400,~450~$MeV, in order to systematically study how they influence the three constraints, i.e., the $2~M_\odot$, the tidal deformability, and the direct URCA constraint.

   We note that other nuclear properties, such as the compressibility $K$ or the slope of symmetry energy $L$ are obtained as in the original parity doublet model without quarks~\cite{Zschiesche:2006zj, Motohiro:2015taa}. For instance, for $m_0=790~$MeV, the hybrid QMN model yields $K=511~$MeV and $L=78~$MeV at the saturation density. Such high value of $K$ is commonly obtained in other field-theoretical models for nuclear matter, such as the linear Walecka model~\cite{Serot:1984ey}. This shortcoming can be accounter for in the context of parity doubling, and one can decrease it to its empirical value $K=240 \pm 20~$MeV by introducing a six-point interaction of the scalar quarkonium $\sigma$~\cite{Motohiro:2015taa} or a scalar tetra-quark~\cite{Gallas:2011qp}. The value of the slope of the symmetry energy is also rather large. It, however, stays within the commonly considered range for the parameter~\cite{Oertel:2016bki, Li:2013ola}. The model parameters to be used in this work are summarized in Table~\ref{tab:model_params}. In-medium profiles of the mean fields are obtained by extremizing the thermodynamic potential~(\ref{eq:thermo_pot_iso}).
   
   Several approaches with parity doubling of baryons exist and have been applied to in-medium QCD and neutron stars~\cite{Hatsuda, Sasaki:2010bp, Dexheimer:2008cv, densePD, Gallas:2011qp, Weyrich:2015hha, densePDPoly, Steinheimer:2011ea}. A different class of quark-hadron hybrid models~\cite{densePDPoly, Steinheimer:2011ea} includes Polyakov loop to thermally suppress quarks in hadronic phase. At finite temperature and small chemical potential, the center of SU(3) gauge group, $Z(3)$ symmetry,  may still be useful to guide the physics of deconfinement to some extent. Under the neutron star conditions, however, the explicit breaking of the $Z(3)$ symmetry is no longer soft, so that it is conceptually questionable to start with this symmetry to constraint any effective Lagrangian. Our hybrid approach is superior in this context to the approaches with the Polyakov loop.
   
   In the next section we discuss the influence of the external parameter $\alpha$ on the equation of state in the hybrid QMN model and its impact on the chiral phase transition, under the neutron-star conditions of $\beta$ equilibrium and charge neutrality.
   
\section{Equation of State}
\label{sec:eos}
   
   The composition of neutron star matter requires \mbox{$\beta$ equilibrium}, as well as the charge neutrality condition. To this end, we include electrons and muons as gases of free relativistic particles. In Fig.~\ref{fig:eos}, we show the calculated zero-temperature equations of state in the mean-field approximation with $m_0=790~$MeV, for different values of the $\alpha$ parameter, namely \mbox{$\alpha b_0=350$~MeV} (solid red line), \mbox{$\alpha b_0=370$~MeV} (dashed magenta line), \mbox{$\alpha b_0=400$~MeV} (dotted blue line) and \mbox{$\alpha b_0=450$~MeV} (dash-dotted black line). The value $b_0$ is the vacuum expectation value of the $b$-field. The mixed phases of the chirally broken and restored phases are shown between circles. We stress that the chiral and hadron-to-quark phase transitions are sequential. The latter happen at higher densities and are not shown in the figure.
   
   In all cases, the \mbox{low-density} behavior is similar. In the case of $\alpha b_0=350$~MeV, the chiral phase transition is triggered at roughly $1.82~\rho_0$, with the mixed phase persisting up to $2.46~\rho_0$. Higher values of $\alpha$ parameter yield weaker transitions at higher densities. For $\alpha b_0=370$~MeV, the mixed phase appears between $2.14-2.73~\rho_0$, and for $\alpha b_0=400$~MeV between $2.68-3.11~\rho_0$. On the other hand, for $\alpha b_0=450$~MeV, the transition turns into a crossover at roughly $3.81~\rho_0$. This stays in correspondence to the case of isospin-symmetric matter, where higher value of $\alpha$ weakens the first-order chiral phase transition, which goes through a critical point, and eventually turns into a crossover transition~\cite{Marczenko:2017huu}.
   
   We stress that since the equations of state are derived by extremizing the thermodynamic potential of Eq.~(\ref{eq:thermo_pot_iso}) with respect to the four mean fields, they are thermodynamically consistent and preserve causality, hence $c_s^2 \leq 1$.
   
   \begin{figure}[t!]
   \begin{center}
         \includegraphics[width=1\linewidth]{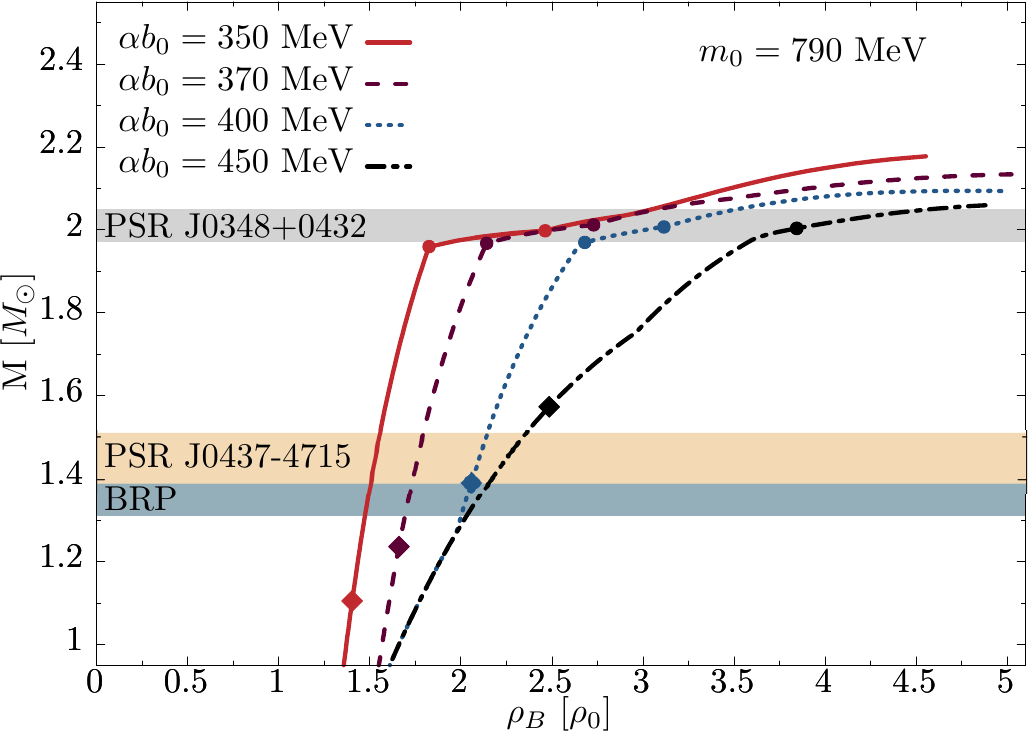}
      \caption{Sequences of masses for compact stars vs. their central net-baryon density as solutions of the TOV equations for $m_0=790~$MeV and four different cases of \mbox{$\alpha b_0=350,~370,~400,~450$~MeV}. The regions between the circles show the coexistence of the chirally broken and chirally restored phases. The diamonds indicate the threshold mass for the direct URCA process (see Sec.~\ref{sec:direct_urca}). The upper gray band is the $2.01(4)~M_\odot$ observational constraint~\cite{Antoniadis:2013pzd}, the orange band, $1.44(7)~M_\odot$, shows the mass of PSR~J0437-4715 for which the NICER experiment~\cite{Reardon:2015kba} will soon provide a radius measurement~\cite{Miller:2016pom}. Finally, the lower blue band is the $1.35(4)~M_\odot$ binary radio pulsar (BRP) constraint~\cite{Thorsett:1998uc}. }
      \label{fig:m_density}
   \end{center}
   \end{figure}
   
   For comparison, in Fig.~\ref{fig:eos-comp}, we show the pressure as a function of the energy density for $\alpha b_0 = 350,~400~$MeV together with the APR EoS~\cite{Akmal:1998cf} (purple dashed line), which is a standard EoS used in astrophysics of compact stars, their mergers, as well as supernovae. The APR EoS shows similar high-density behavior, while being softer in the low-density part, just above the saturation density. Also shown are two EoS taken from Ref.~\cite{Paschalidis:2017qmb}, the multi-polytrope EoS ACB4 (black dash-dotted line), which features a strong first-order phase transition and produces \mbox{high-mass} twin star configurations, as well as the hybrid EoS ACS-1 (green dashed-doubly-dotted line) with a phase transition to constant-speed-of-sound matter at high density.
   
   \begin{figure*}[t]
   \begin{center}
      \includegraphics[width=0.49\linewidth]{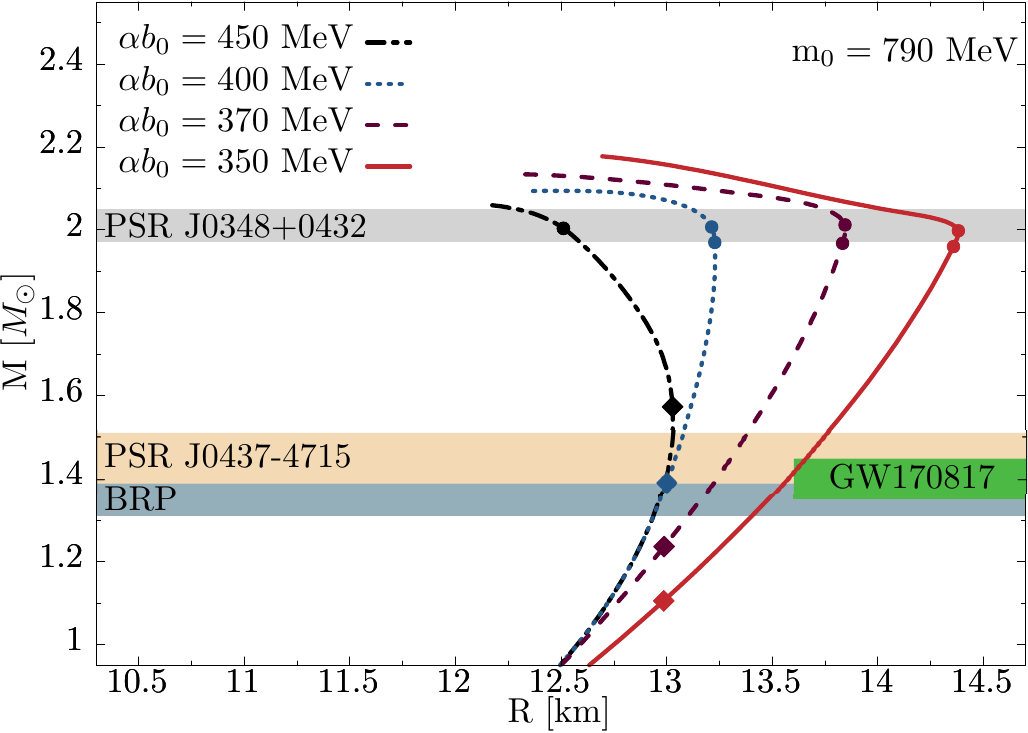}
      \includegraphics[width=0.49\linewidth]{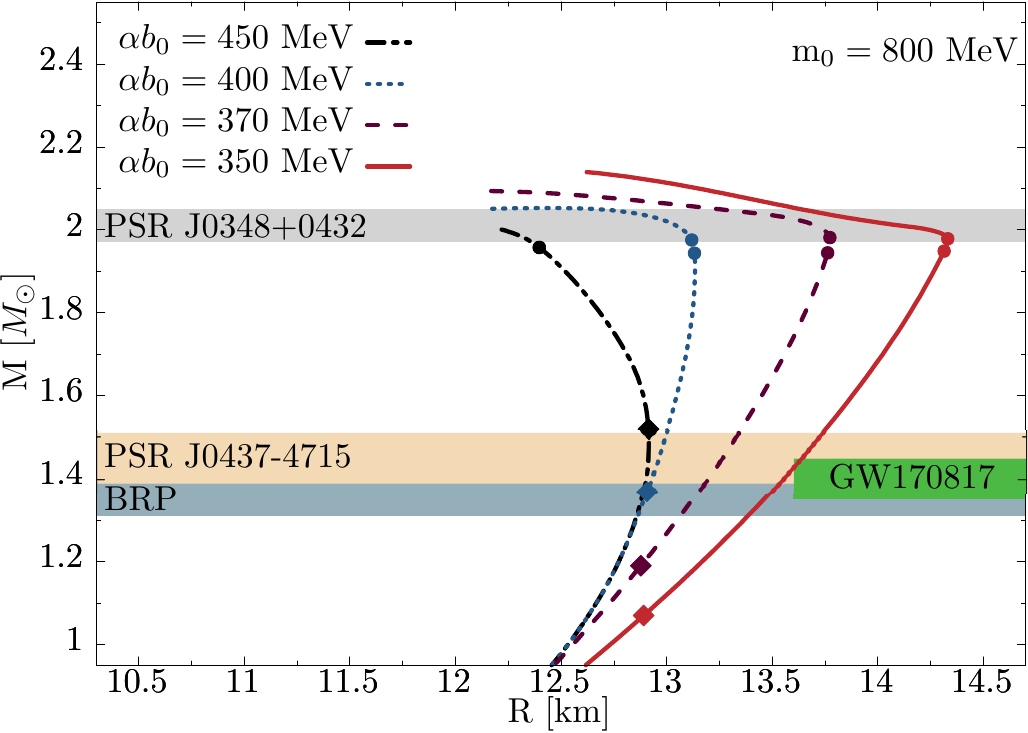}\\
      \includegraphics[width=0.49\linewidth]{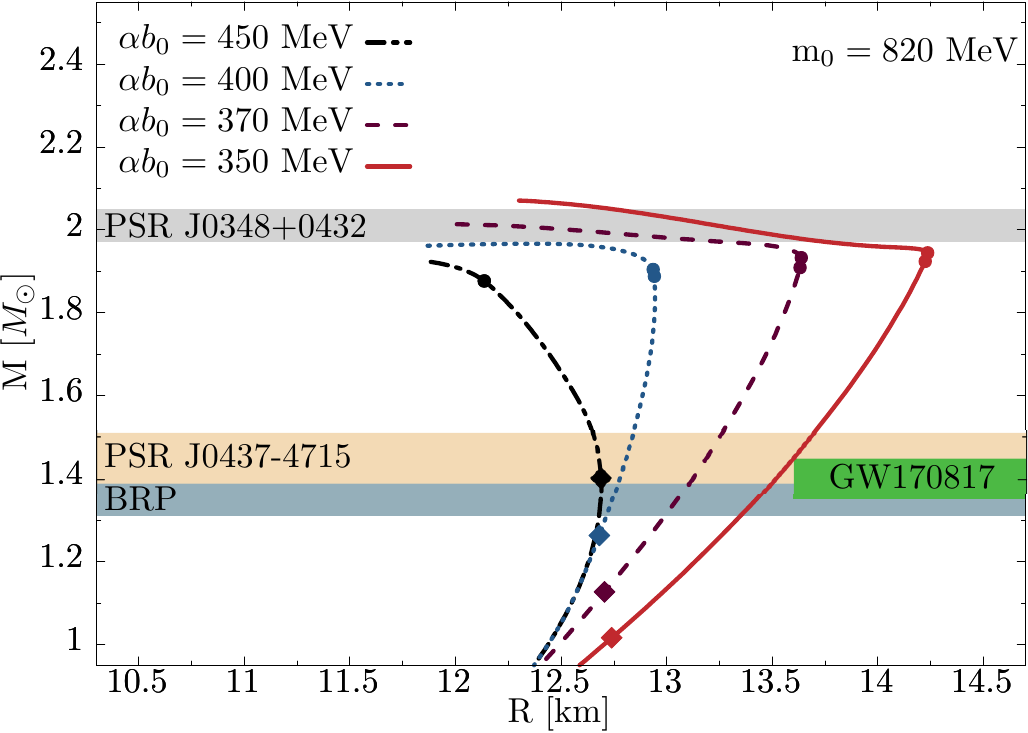}
      \includegraphics[width=0.49\linewidth]{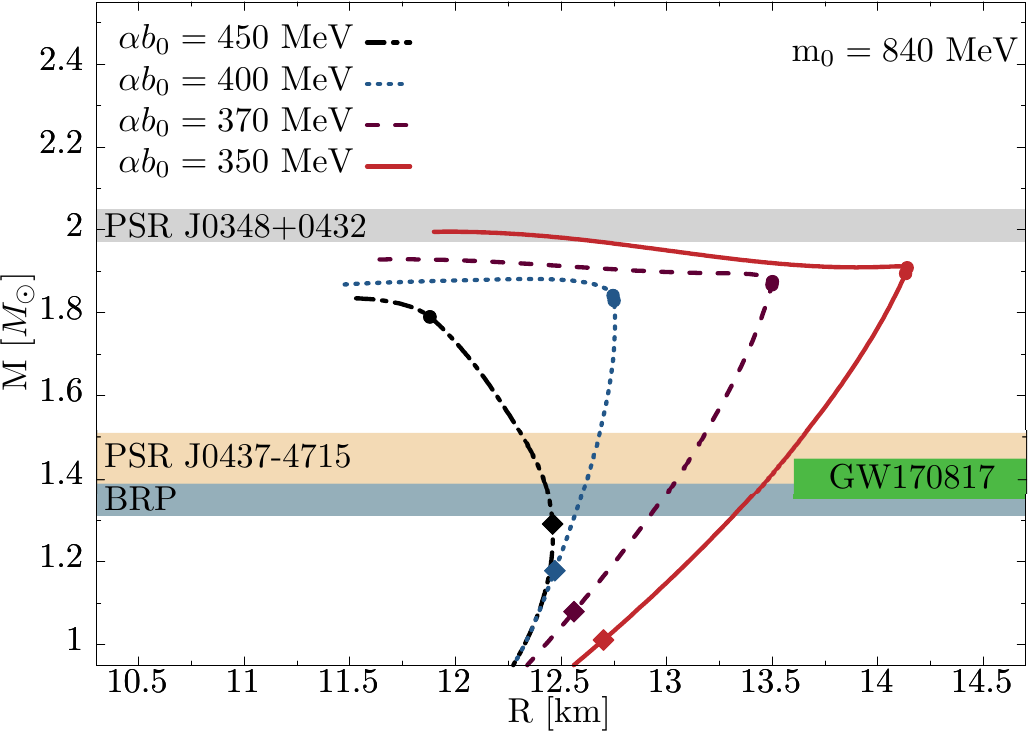}
      \caption{Mass-radius sequences for $m_0=790$~MeV (upper left), $m_0=800$~MeV (upper right), $m_0=820$~MeV (lower left) and $m_0=840$~MeV (lower right), where the legend of the curves as well as the meaning of the circles and diamonds is the same as in Fig.~\ref{fig:m_density}. The green areas "GW170817" show excluded radii of $1.4~M_\odot$ stars extracted in Ref.~\cite{Annala:2017llu}.}
      \label{fig:m_r}
   \end{center}
   \end{figure*}
   
   All the EoS shown in Fig.~\ref{fig:eos-comp} fall into the region derived in~\cite{Hebeler:2013nza} from the maximum mass constraint \mbox{$M_{\rm max}\ge 2.01(4)~M_\odot$}~\cite{Antoniadis:2013pzd} using a multi-polytrope ansatz for the EoS at supersaturation densities (see also~\cite{Alvarez-Castillo:2017qki}). We like to note that the stiffening effect of nuclear matter just above the saturation density when compared to the non-relativistic APR EoS is a feature which the present approach has in common with the relativistic density-functional models of nuclear matter underlying the ACB~\cite{Alvarez-Castillo:2017qki} and ACS~\cite{Alford:2017qgh} class of EoS. The extreme stiffening in the case of $\alpha b_0=350$~MeV and for ACB4 (here due to a nucleonic excluded volume) is at certain tension with the recent analysis of GW170817 by Annala {\it et al}~\cite{Annala:2017llu}. This tension could be resolved, e.g., by a strong phase transition occurring in the compact star mass range relevant for GW170817 (see Ref.~\cite{Paschalidis:2017qmb}).
   
\section{Results}
\label{sec:results}
   
   \subsection{TOV solutions for compact star sequences}
   \label{sec:mass_radius}
      
      We use the equations of state introduced in the previous section (see Fig.~\ref{fig:eos}) to solve the general-relativistic \mbox{Tolman--Oppenheimer--Volkoff} (TOV) equations~\cite{Tolman:1939jz, Oppenheimer:1939ne} for spherically symmetric objects, 
      \begin{subequations}\label{eq:TOV_eqs}
      \begin{align}
         \frac{\dd P(r)}{\dd r} &= -\frac{\left(\epsilon(r) + P(r)\right)\left(M(r) + 4\pi r^3 P(r)\right)}{r \left(r-2M(r)\right)} \textrm,\\
         \frac{\dd M(r)}{\dd r} &= 4\pi r^2 \epsilon(r)\textrm,
      \end{align}
      \end{subequations}
      with the boundary conditions \mbox{$P(r=R) = 0$}, \mbox{$M = M(r=R)$}, where $R$ and $M$ are the radius and the mass of a neutron star, respectively. Once the initial conditions are specified based on a given equation of state, namely the central pressure $P_c$ and the central energy density $\epsilon_c$, the internal profile of a neutron star can be calculated.
      
      In Fig.~\ref{fig:m_density}, we show the relationship of mass versus central net-baryon density, for the calculated sequences of compact stars, together with the state--of--the--art constraint on the maximum mass for the pulsar PSR~J0348-0432~\cite{Antoniadis:2013pzd}. We like to point out that the chiral restoration transition leads to a softening of the EoS so that it is accompanied by a strong increase of the central densities, while the mass of the star is almost unchanged.
      
      \begin{figure}[t!]
      \begin{center}
         \includegraphics[width=1\linewidth]{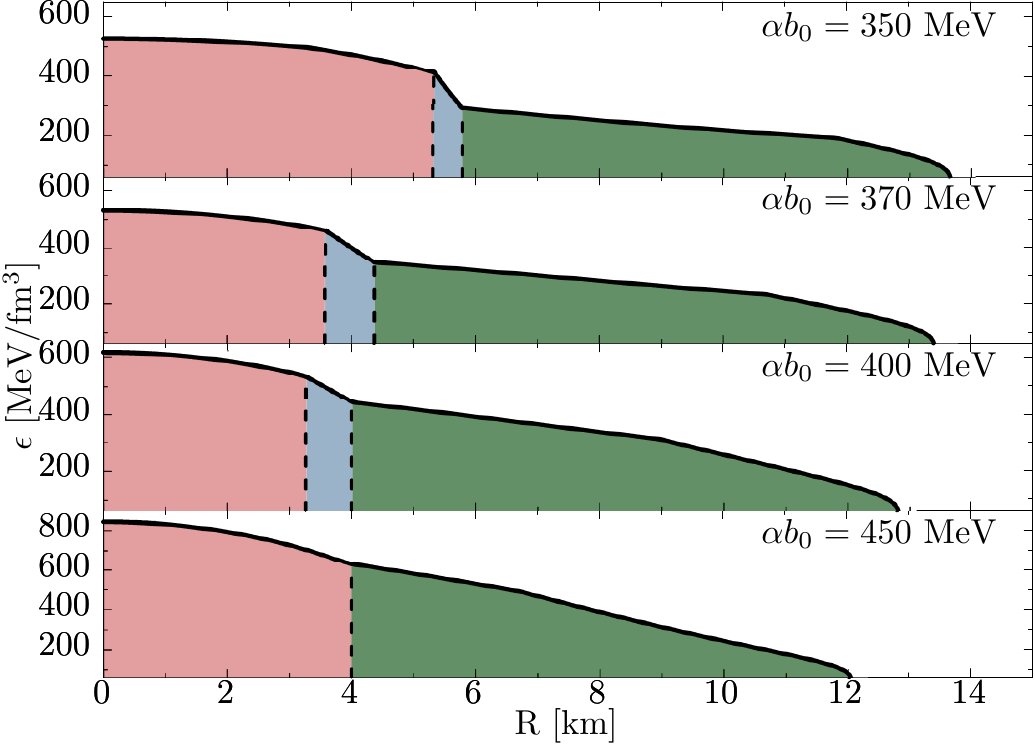}\\
         \includegraphics[width=1\linewidth]{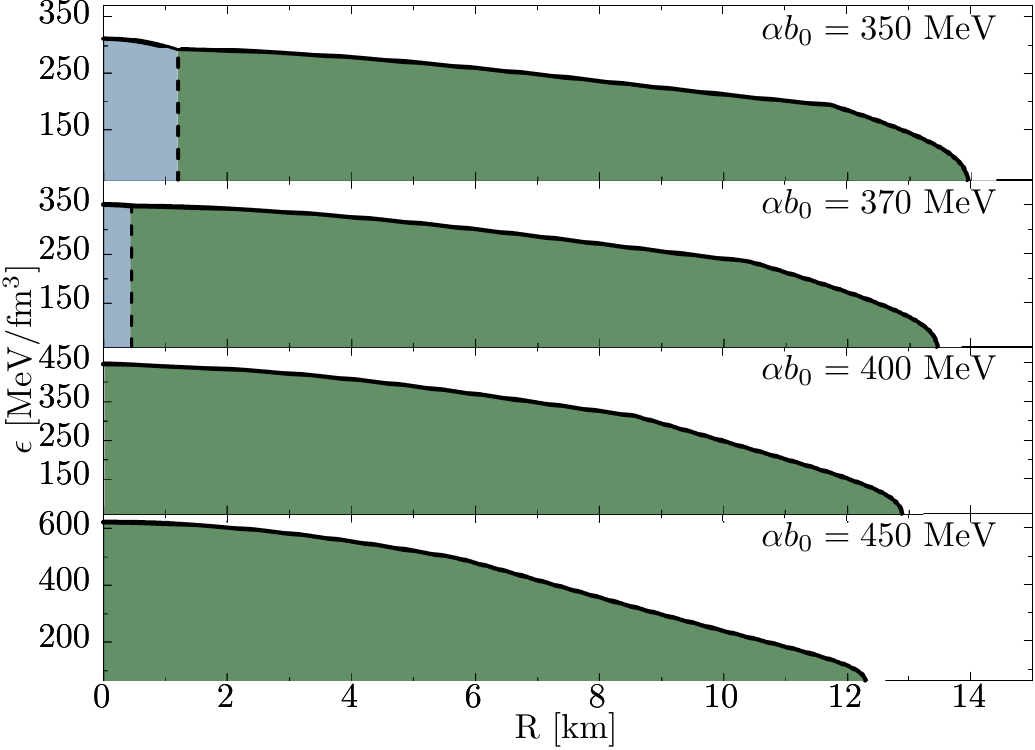}
         \caption{Profiles of the energy density for neutron stars with \mbox{$M = 1.97~M_\odot$} (bottom panel) and \mbox{$M = 2.05~M_\odot$} (top panel) for \mbox{$m_0=790~$MeV}, for four different cases \mbox{$\alpha b_0=350,~370,~400,~450$~MeV}. The green regions show the phase, where the chiral symmetry is broken, in the red regions chiral symmetry is restored, whereas the blue regions indicate the regions of coexistence of both phases (mixed phase).}
         \label{fig:profiles}
      \end{center}
      \end{figure}
      
      In general, there is one--to--one correspondence between an EoS and the \mbox{mass-radius} relation calculated with it.
      The three curves for \mbox{$\alpha b_0 = 350,~370,~400~$MeV} consist of three phases; the chirally broken phase in the low-mass part of the sequence, the chirally restored phase in the \mbox{high-mass} part, and the mixed phase between filled circles. Similarly to the equation of state, increasing the value of $\alpha$ softens the chiral transition, which eventually becomes a smooth crossover for $\alpha b_0 = 450~$MeV and consists only of branches with chiral symmetry being broken and restored, separated by a circle.
      
      We note that the end points of the lines correspond to the onset of quark degrees of freedom in each case, after which the equation of state is not stiff enough to sustain the gravitational collapse and the branches become immediately unstable. This is because, in the current model setup, quarks are not coupled with the vector field leading to a repulsive force. On the other hand, it is known that repulsive interactions tend to stiffen the equation of state. Hence, an additional repulsive force in the quark sector could possibly make the branch stiff enough, so that an additional family of stable hybrid compact stars would appear, with the possibility for the \mbox{high-mass} twin scenario advocated by other effective models~\cite{Alvarez-Castillo:2017qki, Ayriyan:2017nby, Kaltenborn:2017hus}. We leave a further study of quark matter as our future work.
      
      Notably, the chiral transition for all values of $\alpha b_0$ occurs in the \mbox{high-mass} part of the sequence, namely in the vicinity of the \mbox{$2~M_\odot$} constraint, followed by a rapid flattening of the \mbox{mass-radius} sequence. The transitions are, however, not strong enough to produce an additional family of solutions, disconnected by an unstable branch.
      
      In the upper left panel of Fig.~\ref{fig:m_r}, we show the \mbox{mass-radius} relations for the case $m_0=790~$MeV. In the three remaining panels, we show \mbox{mass-radius} relations obtained for different values of the chirally invariant mass $m_0$. What is evident is that increasing the value of $m_0$ systematically strengthens the chiral phase transition. This is seen twofold, as a shrinking of the mixed phases, as well as more abrupt flattening of chirally restored branches. For a larger $m_0$ the transition becomes strong enough to produce disconnected branches (see, e.g., the red solid line in the bottom right panel of Fig.~\ref{fig:m_r}). These, in turn, cause the maximal mass of the \mbox{mass-radius} sequences to decrease with increasing value of $m_0$. Eventually, the equations of state become not stiff enough to reach the $2~M_\odot$ constraint.
      
      \begin{figure*}[t!]
      \begin{center}
         \includegraphics[width=0.49\linewidth]{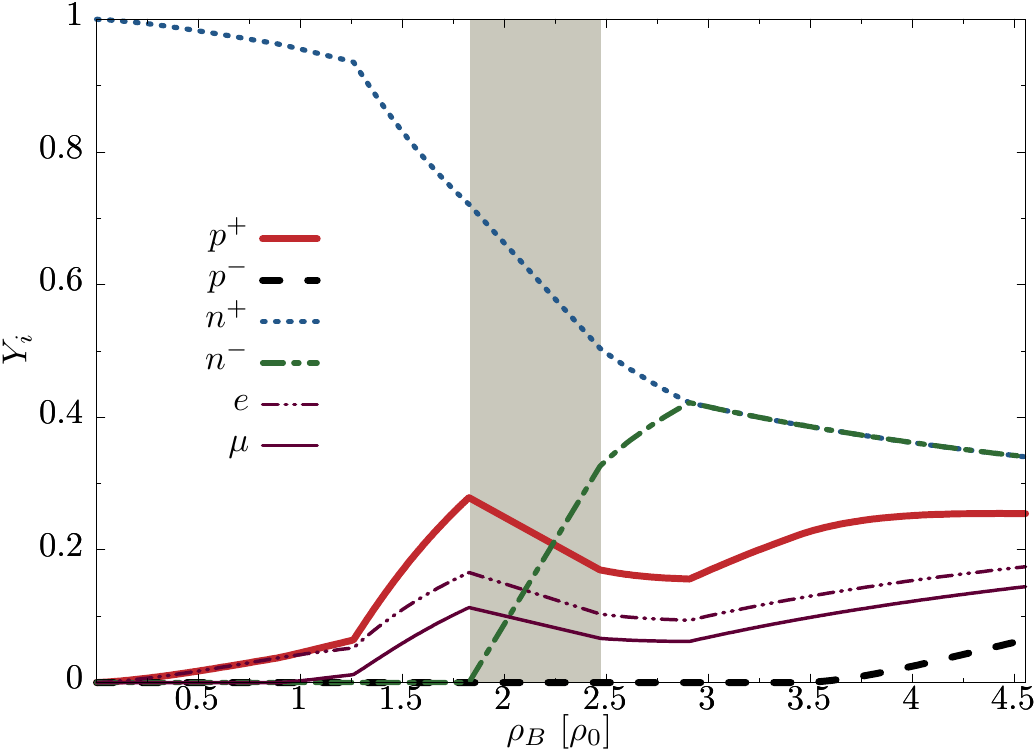}
         \includegraphics[width=0.49\linewidth]{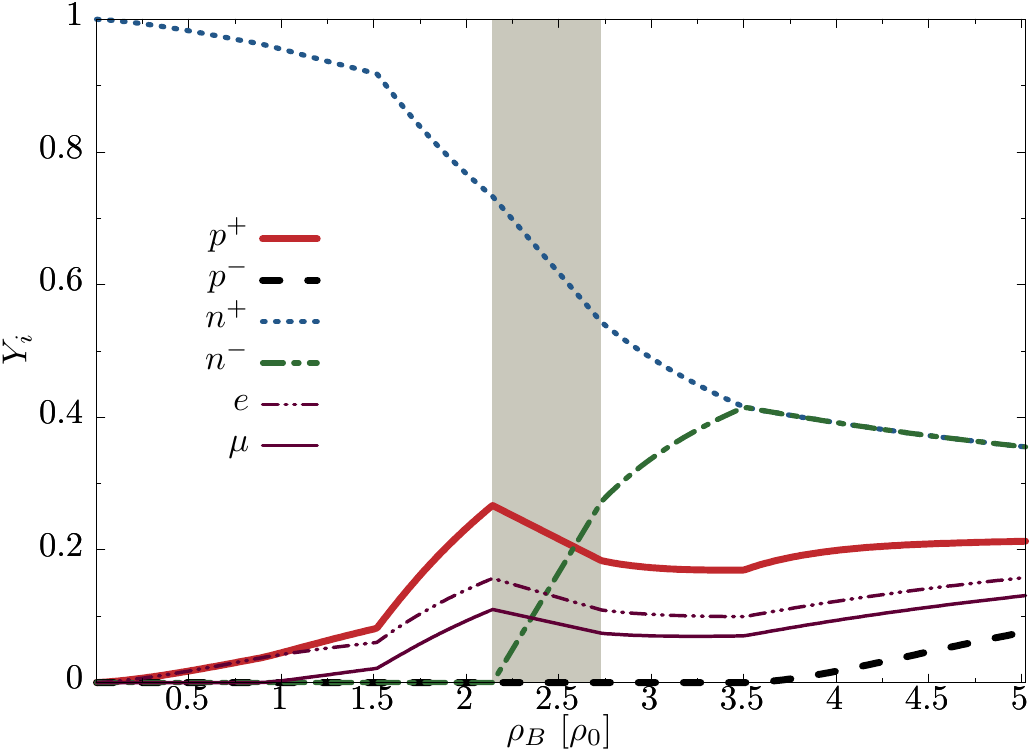}\\
         \includegraphics[width=0.49\linewidth]{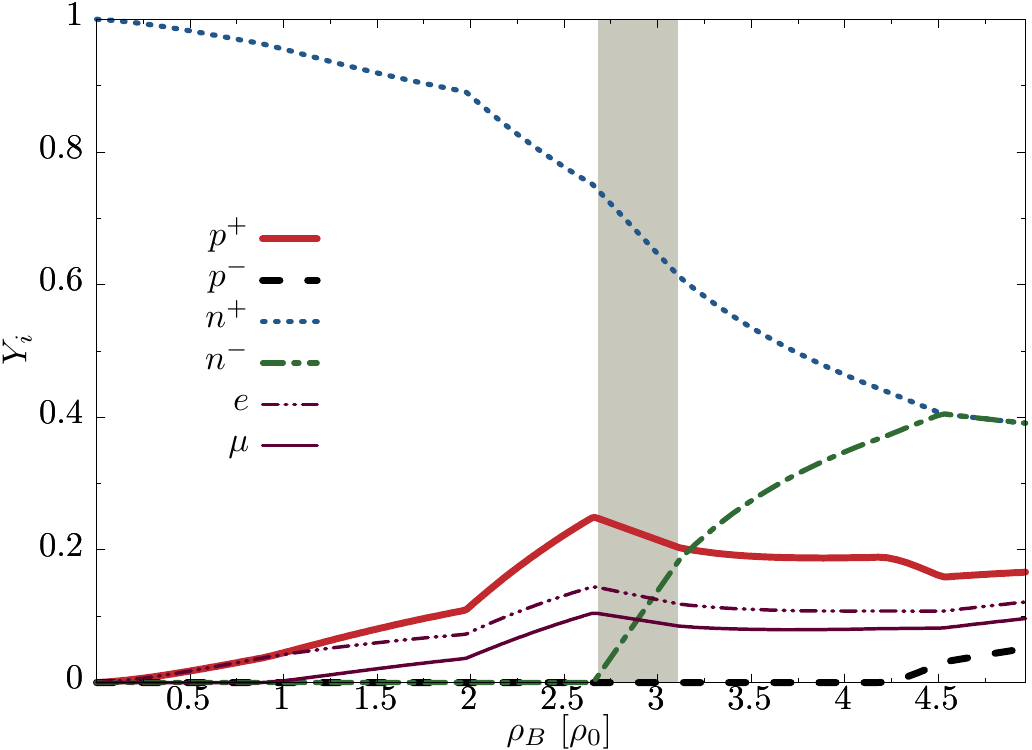}
         \includegraphics[width=0.49\linewidth]{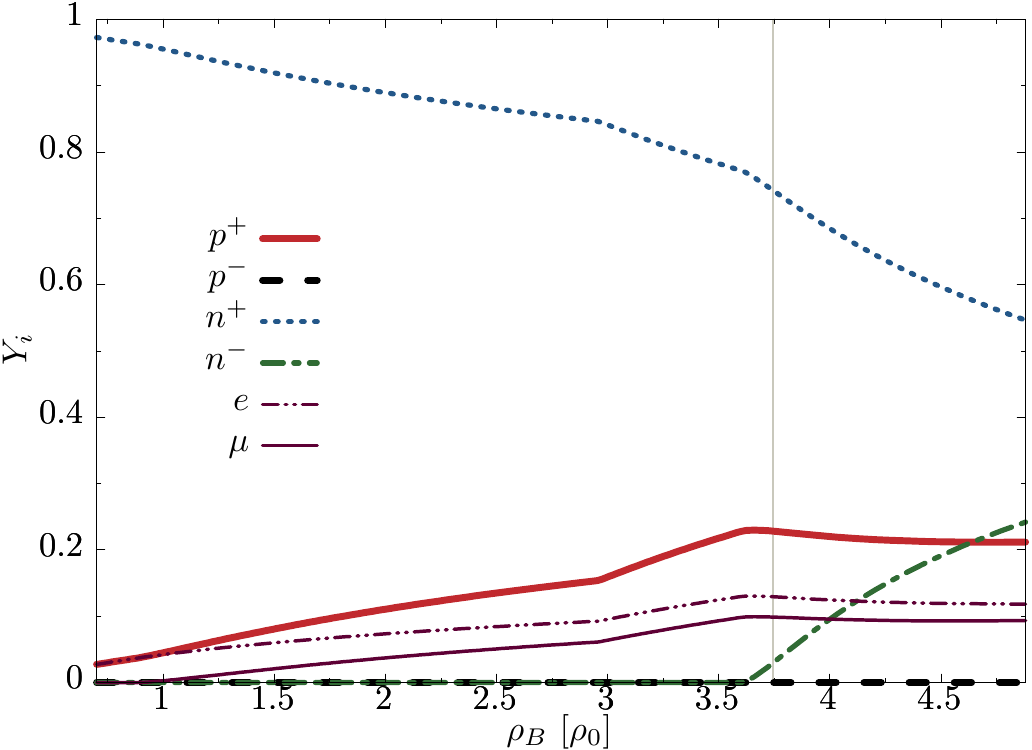}
      \caption{Composition of compact star matter as a function of the baryon density for the present model. The chirally invariant baryon mass is fixed to $m_0=790$~MeV, while four different cases are considered for $\alpha b_0=350$~MeV (upper left), $\alpha b_0=370$~MeV (upper right), $\alpha b_0=400$~MeV (lower left), and $\alpha b_0=450$~MeV~(lower right). The gray shaded area indicate the density regions of coexistence of the broken and restored phases of chiral symmetry.}
      \label{fig:composition}
      \end{center}
      \end{figure*}
      
      We note that the obtained \mbox{mass-radius} relations stay in good agreement with the low-mass constraints derived from the recent neutron-star merger, namely that the radius of a $1.6~M_\odot$ neutron star must be larger than $10.68^{+0.15}_{-0.04}$~km~\cite{Bauswein:2017vtn}. Also, the recently extracted constraint on the radius of $1.4~M_\odot$ stars, that is \mbox{$12.0$~km$~<R(1.4~M_\odot) < 13.6~$km~\cite{TheLIGOScientific:2017qsa, Annala:2017llu, Tews:2018iwm, Most:2018hfd}}, is well fulfilled by all of the parametrizations. In Fig.~\ref{fig:m_r}, the excluded upper bound of this constraint is shown as the green region. Since the radii at $1.4~M_\odot$ are well above $12.0~$km, we do not plot the lower bound of the constraint for the sake of clarity of the figure. From this, one sees that values of $\alpha b_0$ lower than roughly $350~$MeV are excluded by the maximal radius constraint.
      
      In Fig.~\ref{fig:profiles}, we show the energy density profiles of stars for $m_0=790~$MeV. The bottom panel shows profiles of stars with mass $M = 1.97~M_\odot$. The cases with $\alpha b_0=400$ and $450~$MeV show the profiles, which interiors consist only of nuclear matter with broken chiral symmetry (green region). For the stars with $\alpha b_0=350$ and $370~$MeV, the chiral transitions are triggered at roughly $1.2$ and $0.4$~km radii, respectively, and, in their cores only the mixed phase (blue region) is realized. The radii of these stars vary from $13.7$~km to $14.3$~km. In the top panel, we show profiles of stars with mass $M=2.05~M_\odot$. The chiral transitions are triggered in all four cases, and the chirally restored phase is reached in their cores (red region). For $\alpha b_0=350~$MeV, the mixed phase is featured between $5.3-5.8$~km radius, for $\alpha b_0=370~$MeV between $3.6-4.4$~km, for $\alpha b_0=400~$MeV between $3.3-4.1$~km. For $\alpha b_0=450~$MeV, as discussed in previous sections, the transition is a smooth crossover. In this case, we assume the chirally broken and chirally restored phases are separated by the peak in $\partial \sigma / \partial \mu_B$, which happens around $4.0~$km radius. The radii of these stars vary from $12.2$~km to $14.0$~km.
      
      In Fig.~\ref{fig:composition} we show the composition of compact star matter as a function of the baryon density for the present model. The chirally invariant mass of baryons is fixed to $m_0=790~$MeV, while four different cases are considered for \mbox{$\alpha b_0=350~$MeV}~(upper left), \mbox{$\alpha b_0=370~$MeV}~(upper right), \mbox{$\alpha b_0=400$~MeV}~(lower left) and \mbox{$\alpha b_0=450~$MeV}~(lower right). The gray shaded areas indicate the density region of coexistence of the broken and restored phases of chiral symmetry. It is evident that the mixed phase shrinks and the chiral phase transition weakens for higher values of the $\alpha$ parameter.
      
   \subsection{Direct URCA process}
   \label{sec:direct_urca}
      
      The direct URCA (DU) process, $n_+ \rightarrow p_+ + e + \bar \nu_e$, is essential for the cooling of neutron stars and is not expected to occur in neutron stars with masses of the order of $1-1.5~M_\odot$~\cite{Klahn:2006ir}. When triggered, it leads to a substantial enhancement of the neutrino emission, and hence to the neutron star cooling rates. The DU process becomes operative when a critical value of the proton fraction is exceeded~\cite{Lattimer:1991ib}. Taking into account the presence of the parity doublers, as well as assuming that below the deconfinement transition quarks are not populated, i.e. $\rho_B^u=\rho_B^d=0$, the proton fraction is given by
      \begin{equation}\label{eq:p_frac}
         Y_{p_+} = \frac{\rho^{p_+}_B}{\rho_B} = \frac{\rho^{p_+}_B}{\rho^{p_+}_B+\rho^{p_-}_B+\rho^{n_+}_B+\rho^{n_-}_B}\textrm.
      \end{equation}
      In general, $Y_{p_+}$ can be estimated through the momentum conservation condition for the DU process, assuming quasi-equilibrium, $f_{n_+} \leq f_{p_+} + f_e$, where $f_x$'s are the Fermi momenta of neutron, proton and electron, respectively~\cite{Klahn:2006ir}.
      
      \begin{table}[t!]
      \begin{center}
      \begin{tabular}{|c||c|c||c|c|c|c|c|c|c|}
         \hline
         \multicolumn{3}{|c||}{\multirow{3}{*}{}} & \multicolumn{4}{|c|}{$m_0$~[MeV]} \\ \cline{4-7}
         \multicolumn{3}{|c||}{} & $790$
                                 & $800$
                                 & $820$
                                 & $840$ \\ \hline
         $\alpha b_0$~[MeV] & $Y_{p_+}$ & $\rho_B$ $[\rho_0]$ & \multicolumn{4}{c|}{$M$ [$M_\odot$]}\\
         \hline
         \hline
         350 & 13.0\% & 1.41 & 1.12 & 1.09 & 1.04 & 1.01 \\\hline
         370 & 13.2\% & 1.66 & 1.24 & 1.20 & 1.14 & 1.08 \\\hline
         400 & 13.4\% & 2.08 & 1.41 & 1.38 & 1.27 & 1.16 \\\hline
         450 & 13.5\% & 2.51 & 1.58 & 1.53 & 1.41 & 1.29 \\
         \hline
      \end{tabular}
      \end{center}
      \caption{The critical values of the proton fraction ($Y_{p_+}$) for the direct URCA process and corresponding net-baryon number densities ($\rho_B$) and the neutron star masses ($M$).}\label{tab:urca}
      \end{table}
      
      The critical values of masses for the DU process are marked on the \mbox{mass-radius} profiles in Fig.~\ref{fig:m_r} as diamonds. Also shown is the binary-radio-pulsar mass region \mbox{$M=1.35(4)~M_\odot$}~\cite{Thorsett:1998uc} (lower blue band), which sets a lower-bound constraint for the DU threshold in neutron stars. From the figure, it is clear that the DU process becomes operative still in the chirally broken phase, irrespective of the choice of the $m_0$ and $\alpha$ parameters.
      
\begin{figure}[!t]
      \begin{center}
         \includegraphics[width=\linewidth]{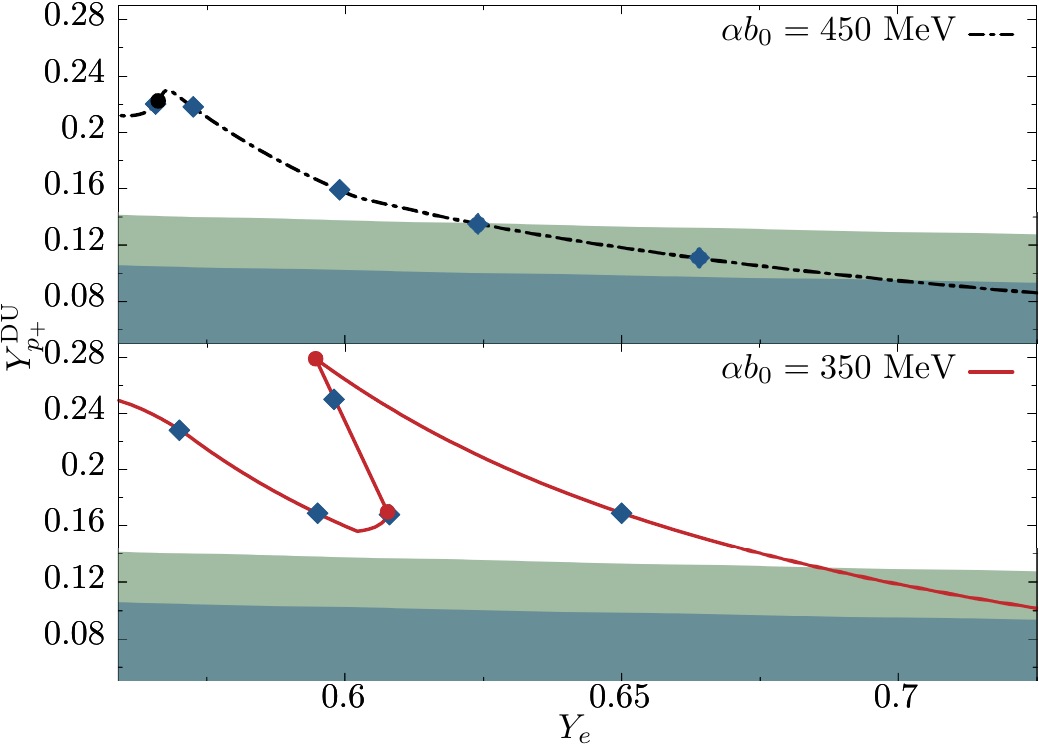}
         \caption{Proton fraction vs. electron fraction under $\beta-$equilibrium and charge neutrality in the hybrid QMN model for \mbox{$m_0=790$~MeV} and two extreme values of \mbox{$\alpha b_0=450$~MeV}~(upper panel) and \mbox{$\alpha b_0=350$~MeV}~(lower panel). The circles indicate the onset and the restoration of chiral symmetry. The blue diamonds on the lines indicate densities $\rho_B/\rho_0=1.5,~2.0,~2.5,~3.0$ and $3.5$ (from the RHS), while the filled circles indicate the onset and end of the chiral restoration transition. The upper gray band shows the region where the direct URCA process is excluded in the case with broken chiral symmetry, while the lower blue band in the case with restored chiral symmetry (see text).}
         \label{fig:y_p}
      \end{center}
      \end{figure}

      Before the chiral transition takes place, the parity partners of proton and neutron are not populated, i.e., their densities are zero. The only relevant degrees of freedom are the positive-parity groundstate nucleons and leptons. Hence, in the chirally broken phase, the situation is similar to the case of ordinary nuclear matter. In that case, the critical value for the proton fraction can be deduced~\cite{Klahn:2006ir},
      \begin{equation}\label{eq:yp1}
         Y^{\rm DU}_{p_+} = \frac{1}{1 + \left(1 + \sqrt[3]{Y_e} \right)^3} \textrm,
      \end{equation}
      where $Y_e = \rho_e / (\rho_e + \rho_\mu)$. The $\rho_e$ and $\rho_\mu$ are the electron and muon densities. $Y_e$ may vary from $1/2$ ($\rho_e = \rho_\mu$) to 1 ($\rho_\mu = 0$). We indicate this range in Fig.~\ref{fig:y_p} (green bands). For $Y_e=1/2$, the critical value is $14.8\%$, and it goes down to $11.1\%$ for the muon-free case.
      
      The above estimate changes when the negative-parity chiral partners are populated. To see this, let us assume a phase with fully restored chiral symmetry. In this limit, we expect that the parity doublers are degenerate, hence their densities are equal. The charge neutrality condition becomes $2\rho_{p_+} = \rho_e + \rho_\mu$, while the momentum conservation condition remains the same~\cite{Lattimer:1991ib}, because only the positive parity states take part in the DU process. This might be modified due to the fact that neutrinos have a finite rest mass which results in a mixing of the left and right handed neutrino sectors. In the present work we are not going to elaborate on this interesting beyond-the-standard-model aspect, for which to best of our knowledge at present no investigation exists in the literature. With this, one finds the threshold to be equal
      \begin{equation}\label{eq:yp2}
         Y^{\rm DU}_{p_+} = \frac{1}{1 + \left(1 + \sqrt[3]{2Y_e} \right)^3} \textrm.
      \end{equation}
      We show this dependence in Fig.~\ref{fig:y_p} (blue bands). For $Y_e=1/2$, the critical value is $11.1\%$, and it goes down to $8.0\%$ for the muon-free case. Note that this estimate is systematically lower than the one from Eq.~(\ref{eq:yp1}).
      
      Also, in Fig.~\ref{fig:y_p}, we show two cases calculated for \mbox{$m_0=790~$MeV}, namely \mbox{$\alpha b_0=350~$MeV}~(bottom panel) and \mbox{$\alpha b_0=450~$MeV}~(top panel). As expected, the critical proton fraction is reached already in the chirally broken phase. The blue diamonds on the lines indicate net-baryon densities, starting with $1.5~\rho_0$ from the RHS, with a step of $0.5~\rho_0$. From this, it is clear that, while the direct URCA process is always operative in the chirally restored phase at high densities (and thus high star masses), for the case $\alpha b_0=350$~MeV it becomes operative already in the chirally broken phase at rather low densities in stars of the typical mass range, so that this parametrization becomes highly unfavorable.
      
      \begin{figure}[t!]
      \begin{center}
         \includegraphics[width=\linewidth]{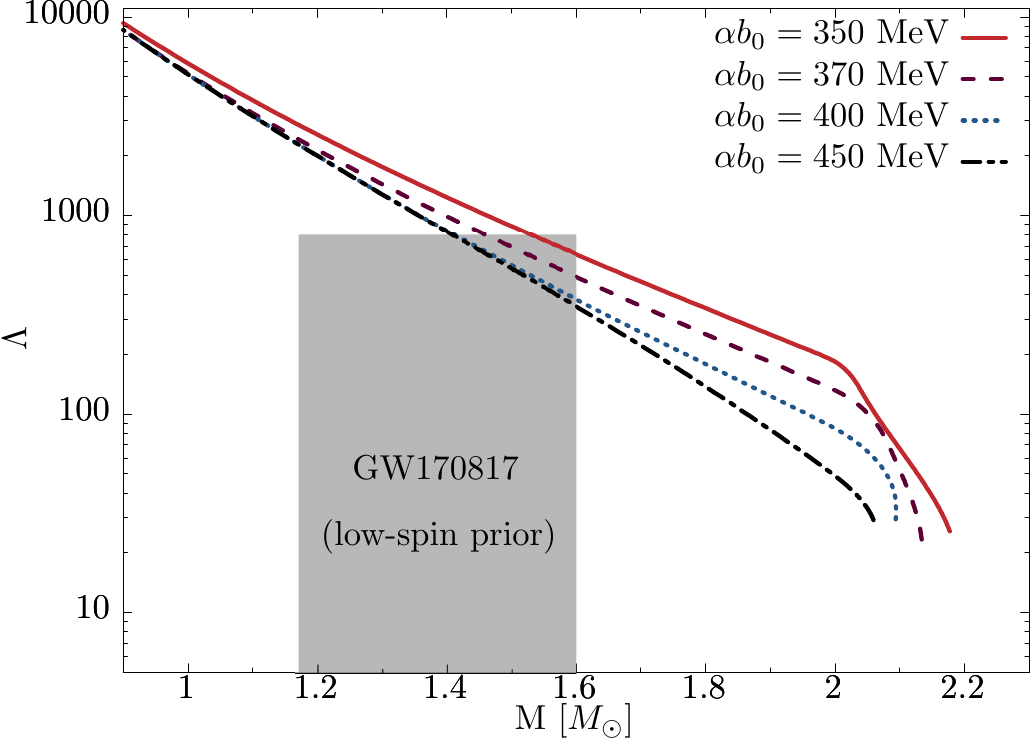}
      \caption{Tidal deformability parameter $\Lambda$ as a function of neutron star mass. The gray box show the $\Lambda<800$ constraint in the range $1.16-1.60~M_\odot$ of the low-spin prior~\cite{TheLIGOScientific:2017qsa}.}
      \label{fig:tidal1}
      \end{center}
      \end{figure}
      
      In Table~\ref{tab:urca}, we show the obtained threshold values of the proton fraction, the corresponding net-baryon number densities and neutron star masses for different values of the chirally invariant mas $m_0$ and parameter $\alpha$. Interestingly, we find that the proton fractions, as well as corresponding net-baryon densities, practically do not depend on the parameter. On the other hand, the neutron star masses, at which the threshold is reached, decrease with increasing $m_0$ and decreasing $\alpha b_0$. This is due to the fact that with lower $\alpha b_0$ the equation of state becomes stiffer, and hence protons are excited more readily at lower values of the net-baryon density. Similarly, the critical value of neutron star mass shifts towards lower values.
      
      While Eq.~(\ref{eq:yp1}) provides a very useful estimate for the DU threshold in the chirally broken phase, the estimate given by Eq.~(\ref{eq:yp2}) should be taken with caution. In the current approach, with two sequential phase transitions, the densities, where the parity doublers are fully degenerate, might be beyond the deconfinement transition. Hence, such scenario would not occur. The situation is similar if the two transitions were simultaneous. In such case, the chiral symmetry restoration is accompanied by the transition from hadrons to quarks.
      
   \subsection{Tidal deformability}
   \label{sec:tidal}
      
      The dimensionless tidal deformability parameter $\Lambda$ can be computed through its relation to the Love number $k_2$~\cite{Hinderer:2007mb,Damour:2009vw,Binnington:2009bb,Yagi:2013awa,Hinderer:2009ca},
      \begin{equation}
         \Lambda = \frac{2}{3} k_2 C^{-5} \textrm,
      \end{equation}
      where $C = M/R$ is the star compactness parameter, with $M$ and $R$ being the total mass and radius of a star. The Love number $k_2$ reads
      \begin{equation}
      \begin{split}
         k_2 &= \frac{8C^5}{5} \left(1-2C\right)^2 \left[ 2+2C(y-1)-y \right] \times \\
            \times & \Big(2C\left[6- 3y + 3C(5y-8) \right]\\
            +      & 4C^3[13-11y+C(3y-2)+2C^2(1+y)] \\
            +      & 3(1-2C)^2[2-y+2C(y-1)\ln{(1-2C)}]\Big)^{-1} \textrm,
      \end{split}
      \end{equation}
      where $y = R\beta(R)/H(R)$. The functions $H(r)$, and $\beta(r)$ are given by the following set of differential equations,
      \begin{eqnarray}
      \frac{\dd \beta}{\dd r}&=&2 \left(1 - 2\frac{M(r)}{r}\right)^{-1} \nonumber\\
      && H\left\{-2\pi \left[5\epsilon(r)+9 P(r)+\frac{\dd \epsilon}{\dd P}(\epsilon(r)+P(r))\right]\phantom{\frac{3}{r^2}} \right. \nonumber\\
      && \left.+\frac{3}{r^2}+2\left(1 - 2\frac{M(r)}{r}\right)^{-1} \left(\frac{M(r)}{r^2}+4\pi r P(r)\right)^2\right\}\nonumber\\
      &&+\frac{2\beta}{r}\left(1 - 2\frac{M(r)}{r}\right)^{-1}\nonumber\\
      &&  \left\{\frac{M(r)}{r}+2\pi r^2 (\epsilon(r)-P(r)) - 1\right\}~\textrm,\\
      \frac{\dd H}{\dd r}&=& \beta \textrm.
      \end{eqnarray}
      The above equations have to be solved along with the TOV equations~(\ref{eq:TOV_eqs}). The initial conditions are \mbox{$H(r\rightarrow 0) = c_0 r^2$} and \mbox{$\beta(r\rightarrow 0) =2c_0 r$}, where $c_0$ is a constant, which is irrelevant in the expression for the Love number $k_2$.
      
      \begin{figure}[t!]
      \begin{center}
         \includegraphics[width=\linewidth]{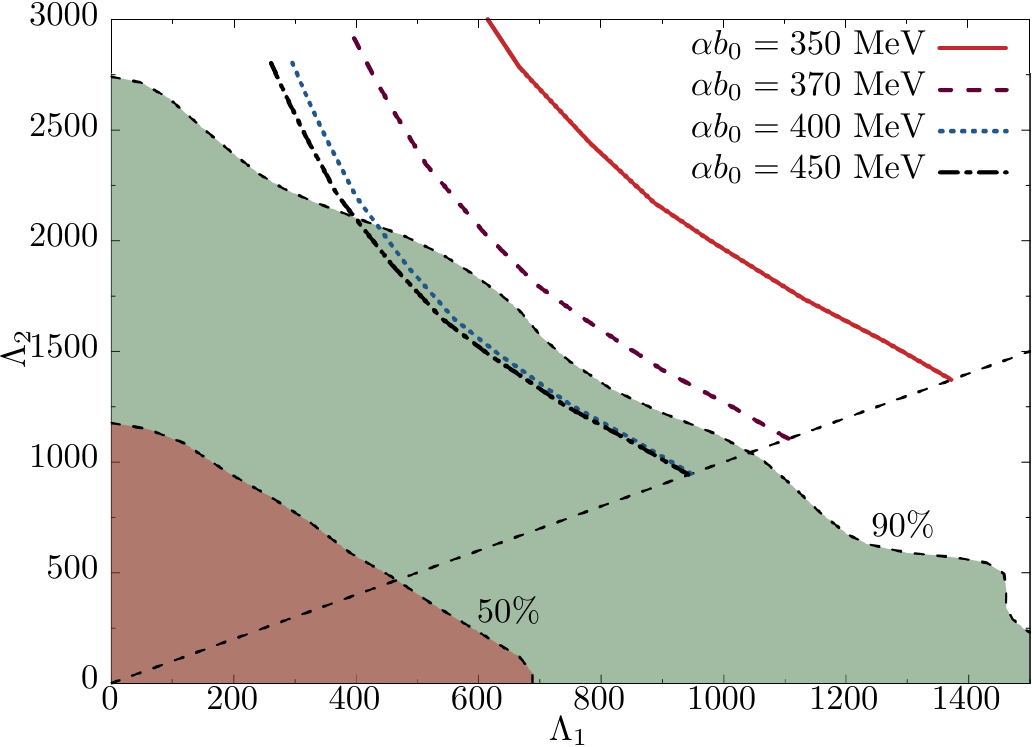}
      \caption{Tidal deformability parameters $\Lambda_1$ and $\Lambda_2$ of the low- and \mbox{high-mass} mergers obtained from $\Lambda(m)$ relation for \mbox{$m_0=790~$MeV}. Shown are also the $50\%$ and $90\%$ probability contours for the low-spin prior~\cite{TheLIGOScientific:2017qsa}.}
      \label{fig:tidal2}
      \end{center}
      \end{figure}
      
      In Fig.~\ref{fig:tidal1}, we show the dimensionless tidal deformability parameter $\Lambda$ as a function of neutron star mass $M$, for \mbox{$m_0=790~$MeV}. We also show the constraint derived in~\cite{TheLIGOScientific:2017qsa}, $\Lambda\left(1.4~M_\odot\right) < 800$. The constraint is met only for the cases with $\alpha b_0 = 400$ and $450~$MeV. In Fig.~\ref{fig:tidal2}, we plot the tidal deformability parameters $\Lambda_1$ vs. $\Lambda_2$ of the high- and low-mass members of the binary merger together with the $50\%$ and $90\%$ fidelity regions obtained by the LVC analysis of the GW170817 merger event~\cite{TheLIGOScientific:2017qsa}. We note that the tidal deformability parameter favors equations of state that are soft around the saturation density. On the other hand, the $2~M_\odot$ constraint requires a sufficiently stiff equation of state at higher densities. The interplay between the two constraints can be further used to fix the model parameters $m_0$ and $\alpha b_0$. We discuss this matter in the next subsection.
      
   \subsection{Isospin-symmetric phase diagram}
   \label{sec:qcd_phase_diagram}
      
      \begin{figure}[t!]
      \begin{center}
         \includegraphics[width=\linewidth]{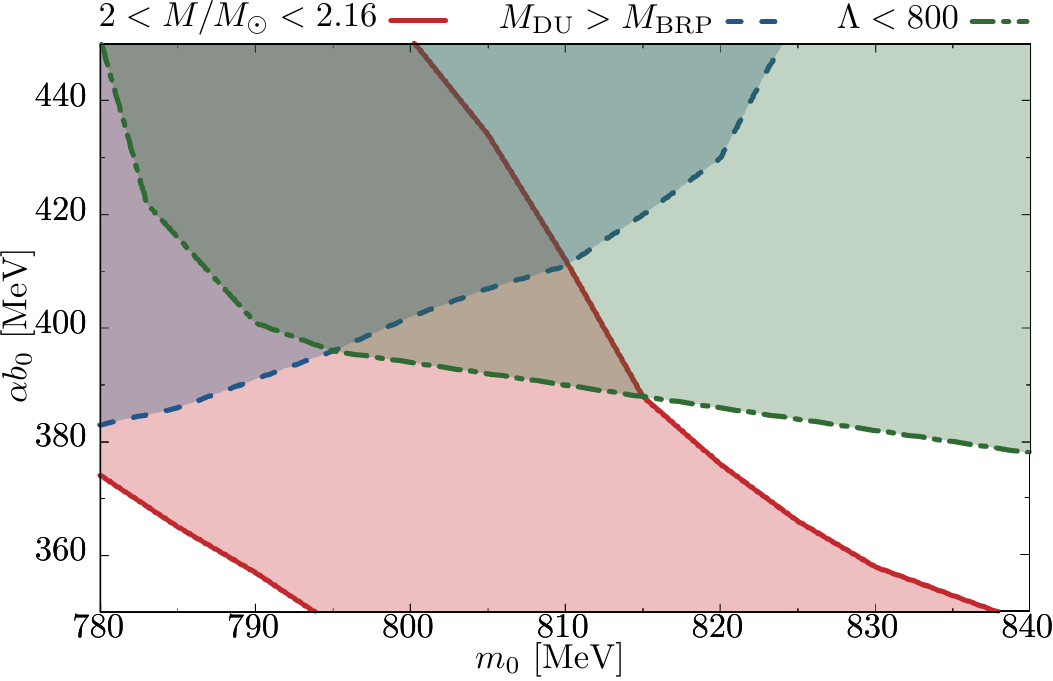}
         \caption{Constraints for the model parameters $m_0$ and $\alpha b_0$: The maximum mass constraint (red solid line); the direct URCA (blue dashed line); the tidal deformability constraint (green broken-dashed line). The corresponding shaded ares show regions where the constraints are met (See text for details).}
         \label{fig:constraints}
      \end{center}
      \end{figure}
      
      The observational neutron-star data provide useful constraints on the structure of strongly interacting matter. Furthermore, they may constrain the phase diagram of isospin-symmetric QCD matter, which is of major relevance for the heavy-ion physics.
      
      \begin{figure}[t!]
      \begin{center}
         \includegraphics[width=\linewidth]{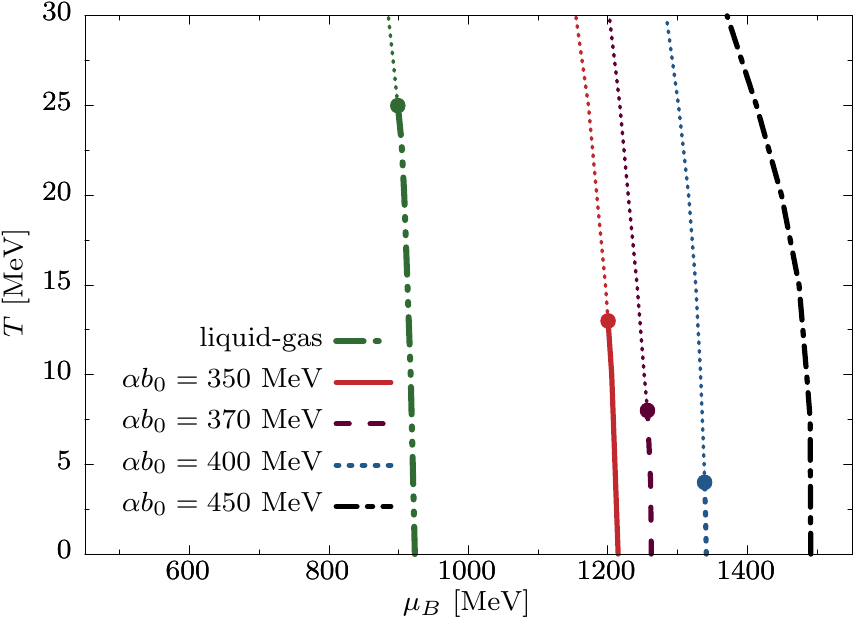}
         \caption{Low-temperature part of the phase diagram in the $(T,\mu_B)$-plane for isospin-symmetric matter obtained in the hybrid QMN model. The green \mbox{dashed-doubly-dotted} curve corresponds to the liquid-gas phase transition common for all $\alpha b_0$. The circles indicate critical points on the transition lines above which, the first-order transition turns into a crossover. No critical point is shown for the case with $\alpha b_0=450$~MeV, for which the chiral transition is a smooth crossover at all temperatures.}
         \label{fig:phase_diag}
      \end{center}
      \end{figure}
      
      In Fig.~\ref{fig:constraints}, we compile the three constraints discussed in this work in the $(\alpha b_0,m_0)$-plane. Firstly, we show the constraint on the maximum mass of non-rotating, cold neutron stars (red solid line), namely
      \begin{equation}
         2.01 < M/M_\odot < 2.16 \textrm.
      \end{equation}
      These limits were was deduced from the assumption that the GW170817 merger event could not have collapsed directly into a black hole, but rather formed a super massive and rapidly-rotating neutron star~\cite{Rezzolla:2017aly,Most:2018hfd}. The lower limit coincides with the mass measurement of PSR~J0348+0432~\cite{Antoniadis:2013pzd}. Similar upper limits on the maximum mass was derived independently in Refs.~\cite{Margalit:2017dij, Shibata:2017xdx}. Secondly, we show the constraint on our parameter space that results from the direct URCA constraint, i.e., from the existence of a lower bound on the mass of neutron stars for which the fast direct URCA process could become operative. We choose this threshold mass rather conservatively to be the upper limit of the mass range of binary radio pulsars, $M_{\rm BRP} \approx 1.4~M_\odot$, representing the upper limit for the masses of "typical" neutron stars. Finally, the constraint  on the tidal deformability parameter, $\Lambda(1.4~M_\odot) < 800$, induced from the GW170817 merger event analysis~\cite{TheLIGOScientific:2017qsa} is shown as the green broken-dashed line. The areas with corresponding shaded colors show regions where the constraints are met. The region where they overlap gives the most preferable sets of the external parameters $\alpha b_0$ and $m_0$. We note that, by construction, the allowed range for $\alpha b_0$ is roughly $300-450~$MeV~\cite{Benic:2015pia,Marczenko:2017huu}. Since higher $m_0$, in general, yields softer equation of state, the stiffness required by the maximum mass constraint is compensated by lower values of $\alpha b_0$, and inversely for the remaining two constraints. At lower values of $m_0$, the maximum mass constraint is met practically for the whole range of $\alpha b_0$, and the region of overlap is set by the other two constraints (top left corner of the figure). Hence, in general higher values of $\alpha b_0$ are more preferable. The three constraints are met down to roughly $m_0=780~$MeV. This sets the lower and upper bounds for the $m_0$ parameter. As a result, the allowed range for $m_0$ is roughly $m_0=780-810~$MeV.
      
      In Fig.~\ref{fig:phase_diag}, we show the phase diagram obtained in the model in the \mbox{$(T,\mu_B)$-plane}, for the case of $m_0=790~$MeV. The first-order liquid-gas phase transition (green \mbox{dashed-doubly-dotted} line) develops a critical point around $T=25~$MeV, and turns into a crossover above it, similarly to the pure parity doublet model. We note that by construction of the hybrid QMN model, the liquid-gas phase transition line is common for all values of the $\alpha$ parameter~\cite{Benic:2015pia, Marczenko:2017huu}. Similar phase structure is seen in the chiral phase transition for $\alpha b_0=350,~370,~400~$MeV, which develop critical points at $T=13,~8,~4~$MeV, respectively. On the other hand, for $\alpha b_0 = 450~$MeV, the chiral transition is a smooth crossover at all temperatures. Note that, in view of the constraints discussed in this work (see Sec.~\ref{sec:results}), the scenarios with $\alpha b_0=350~$MeV and $\alpha b_0=370~$MeV are rather excluded. Hence, either rather low temperature for the critical endpoint or even its absence in the phase diagram for isospin-symmetric matter is favored.
      
\section{Conclusions}
\label{sec:conclusion}
   
   In this work, we investigated the consequences of a recently developed hybrid quark-meson-nucleon (QMN) model for the equation of state of dense matter under neutron star conditions and the phenomenology of compact stars. In particular, we focused on the implications of the realization of the chiral symmetry restoration by parity doubling within the hadronic phase. We have demonstrated that a strong first-order phase transition invalidates the implication that a flattening, eventually even occurrence of a mass-twin phenomenon, of the \mbox{mass-radius} relation for compact stars at $2~M_\odot$, could inevitably signal a deconfinement phase transition in compact stars~\cite{Alvarez-Castillo:2016wqj}. We have explored different scenarios for \mbox{high-mass} neutron stars with masses in the range of $2~M_\odot$. We have shown that, within the hybrid QMN model, the \mbox{high-mass} stars, such as the PSR~J0348+0432 pulsar with the mass $2.01(4)~M_\odot$~\cite{Antoniadis:2013pzd}, can be realized in different ways. First, as a neutron star with single-phase nuclear matter with broken chiral symmetry. Second, where its core is made of a mixed phase surrounded by chirally broken and confined nuclear matter. Finally, as a two-phase nuclear matter star, where the core with chiral symmetry restored but still confined matter is surrounded by a chirally broken and confined phase.
   
   We emphasize that an abrupt change in a \mbox{mass-radius} profile in the \mbox{high-mass} part of the sequence is, in general, a result of a phase transition. As we have shown in this work, in a model with two sequential transitions, it does not need to be associated with the deconfinement transition, and hence does not imply the existence of quark matter in the core of a neutron star. We discussed the dependence of the interior composition on the variation of the model parameter $m_0$, which has implications on the mass of the star for which the direct URCA process sets in, as well as on the compactness of the star. 
   
   The most favorable parametrization is found to be $m_0 \approx 790-800~$MeV and $\alpha b_0 \gtrsim 400~$MeV. For this set of parameters, the chiral phase transition occurs around or even within the $2.01(4)~M_\odot$. Therefore, the resulting $2~M_\odot$ neutron stars with the chirally broken phase in their outer parts, may go through the mixed phase and eventually reach the chirally restored phase in their cores. This is well illustrated in Fig.~\ref{fig:profiles} where the energy-density profiles obtained for $m_0=790~$MeV are shown at the lower and upper bounds of the $2.01(4)~M_\odot$ constraint. The stars at the lower bound do not reach the chirally restored phase in their cores, while the ones at the upper bound do.
   
   We have discussed how modern astrophysical constraints compiled together allow for better determination of the available parameter range, and demonstrated that not only too soft (excluded by the maximum mass constraint), but also too stiff (excluded by either the direct URCA or the tidal deformability constraint) equations of state may be ruled out in the current approach. Finally, we have shown that, due to the parity doubling phenomenon, the obtained results suggest rather low value of the temperature for the critical endpoint of the \mbox{first-order} chiral phase transition in the phase diagram, which eventually may even be absent.
   
   We note that one of the shortcomings of the hybrid QMN model in the current setup is the high value of the compressibility parameter. Improving our model by introducing additional scalar interaction would make the matter more compressible, thus make the equation of state softer around the saturation density. This would have an impact on neutron stars in the low-mass part of the sequence. Hence, one should expect a smaller radius of the corresponding neutron star. Since the tidal deformability scales with the inverse of the compactness parameter, one should also expect a smaller value for the tidal deformability parameter as well. This would make the comparison of our model with the tidal deformability constraint from the GW170817 event better. We plan to elaborate on this our future work.
   
   In view of the recent formulation of the three-flavor parity doubling~\cite{Steinheimer:2011ea, Sasaki:2017glk} and further lattice QCD studies~\cite{Aarts:2015mma, Aarts:2017rrl}, where it was found that to large extent the phenomenon occurs also in the hyperon channels, it would be of great interest to establish equations of state that include these degrees of freedom. In general the inclusion of heavier flavors is known to soften the equation of state and additional repulsive forces are needed to comply with the $2~M_\odot$ constraint. Possible resolutions are strong repulsion between hyperons and nucleons induced at high momenta~\cite{Haidenbauer:2013oca, Yamamoto:2015lwa} or finite eigenvolumes of hadrons~\cite{Mukherjee:2017jzi}. Besides, additional stiffness from the quark side would play a role, which is not included in the current study. We note that it may be essential for establishing the branch of stable hybrid star solutions. Work in this direction is in progress and the results will be reported elsewhere.
   
\acknowledgments
   
   MM acknowledges comments and fruitful discussions with \mbox{N.-U.~F.~Bastian}, \mbox{D.~E.~Alvarez-Castillo}, \mbox{T.~Fischer} and \mbox{M.~Szyma\'nski}. This work was partly supported by the Polish National Science Center (NCN), under Maestro grant no. DEC-2013/10/A/ST2/00106 (KR and CS), Opus grant no. UMO-2014/13/B/ST9/02621 (MM and DB), and Preludium grant no. UMO-2017/27/N/ST2/01973 (MM). DB is grateful for support within the MEPhI Academic Excellence programme under contract no. 02.a03.21.0005. We acknowledge the COST Actions CA15213 "THOR" and CA16214 "PHAROS" for supporting networking activities.

\end{document}